%% file: PREPRINT_Article01TransmissionOfElectriFieldSpineAPSOS_NOFIG.tex
\documentclass[%
 preprint,
 amsmath,amssymb,
 aps,
longbibliography
]{revtex4-2}

\usepackage{graphicx}
\usepackage{subcaption}	
\usepackage{dcolumn}
\usepackage{bm}
\usepackage[hidelinks=True]{hyperref}

\usepackage[french]{babel}  
\usepackage[T1]{fontenc}
\usepackage[utf8]{inputenc} 
\newcommand{\uvec}[1]{\boldsymbol{\hat{\textbf{#1}}}}
\newcommand{\freq}[1]{\text{ {#1}Hz}}
\newcommand{\RR}{\text{R}}
\newcommand{\LL}{\text{L}}
\newcommand{\TT}{\text{T}}
\newcommand{\rmi}{\text{i}}
\newcommand{\rme}{\text{e}}

\begin{document}


\title{Electric Field Propagation with Spinal Cord Stimulation}

\author{O. Bernard$^{1,2}$}
\email{Contact author: Olivier.L.Bernard@Usherbrooke.ca}
\author{R. Fontaine$^{3}$}%
\author{O. Eddaoui$^{4}$}
\author{J. Harnie$^{4}$}
\author{A. Frigon$^{4}$}
\author{C. Iorio-Morin$^{2}$}


\affiliation{
$^{1}$Institut Quantique, Département de Physique, Université de Sherbrooke, Québec, Canada, J1K 2R1
}%
\affiliation{
 $^{2}$Service de Neurochirurgie, Département de Chirurgie, Université de Sherbrooke, Québec, Canada,  J1H 5N4 
}%

\affiliation{%
$^{3}$3IT- Department of electrical engineering and computer engineering, Université de Sherbrooke, Québec, Canada J1K 2R1 
}%

\affiliation{%
$^{4}$Department of Pharmacology- Physiology, Université de Sherbrooke, Québec, Canada, J1H 5N4 
}%


\date{\today}

\begin{abstract}
	Spinal cord stimulation is routinely used for the treatment of chronic pain, and is increasingly being investigated for the restoration of movement after paralysis. However, most of our current knowledge on epidural spinal cord stimulation relies on empirical approaches and semi-hybrid models. Hence, optimizing this therapy requires a better theoretical understanding of how stimulation affects the target neural structures. 
	Using the physical properties of tissues combined with an electromagnetic, dielectric, and electrolytic conductor description of them, we calculated the upper analytical limit of electric field energy transmission in target neural tissues for different frequencies in the spinal cord using a dual-frequency framework: a quasi-static description for low frequencies ($< 1\text{ MHz}$) and a multi-layer transfer-matrix model for high-frequency radiation regime ($>1\text{ MHz}$).  
	 Simulations suggest that the propagation through the tissues of the spinal cord of an epidurally applied electric field has a complex relationship with the field pulse width. In addition the electric field energy is strongly absorbed at the junction between dura mater and cerebrospinal fluid, and within the cerebrospinal fluid as induced current. These currents hit the different bodies floating within the cerebrospinal fluid, and/or dorsal horn in a very localized region of the spinal cord.
	The proposed models allow theoretical calculation of the transmission factor for different numbers of layers of tissue within the spinal canal, and for different electric field pulse widths. This eases the prediction of the expected energy reaching each layer of the spinal cord for different tissues, parameters, and electrode configurations, simplifying the optimization of spinal cord stimulation treatment.
\end{abstract}

\maketitle


	\section{Introduction}
	
	Electric Field (EF) stimulation has long been used to treat various human diseases, and injuries \cite{Ranck1965,Coburn1985}. An alternating electric current is applied using one or more pairs of electrodes positioned on a region of interest. 
	This generates an EF that propagates into underlying tissues. The response varies depending on the target (e.g., muscle, nerve, etc.), and often results in a physiological output that can be used for therapeutic benefit. Examples include pacemakers leading to heart muscle contractions, transcutaneous electrical nerve stimulation (TENS) used for pain relief, and deep brain stimulation (DBS) used in Parkinson’s disease, obsessive-compulsive disorder, or epilepsy \cite{Rogers2022}.
	
\begin{figure}[ht!]
	    \centering
	    \includegraphics[ width=8.5cm]{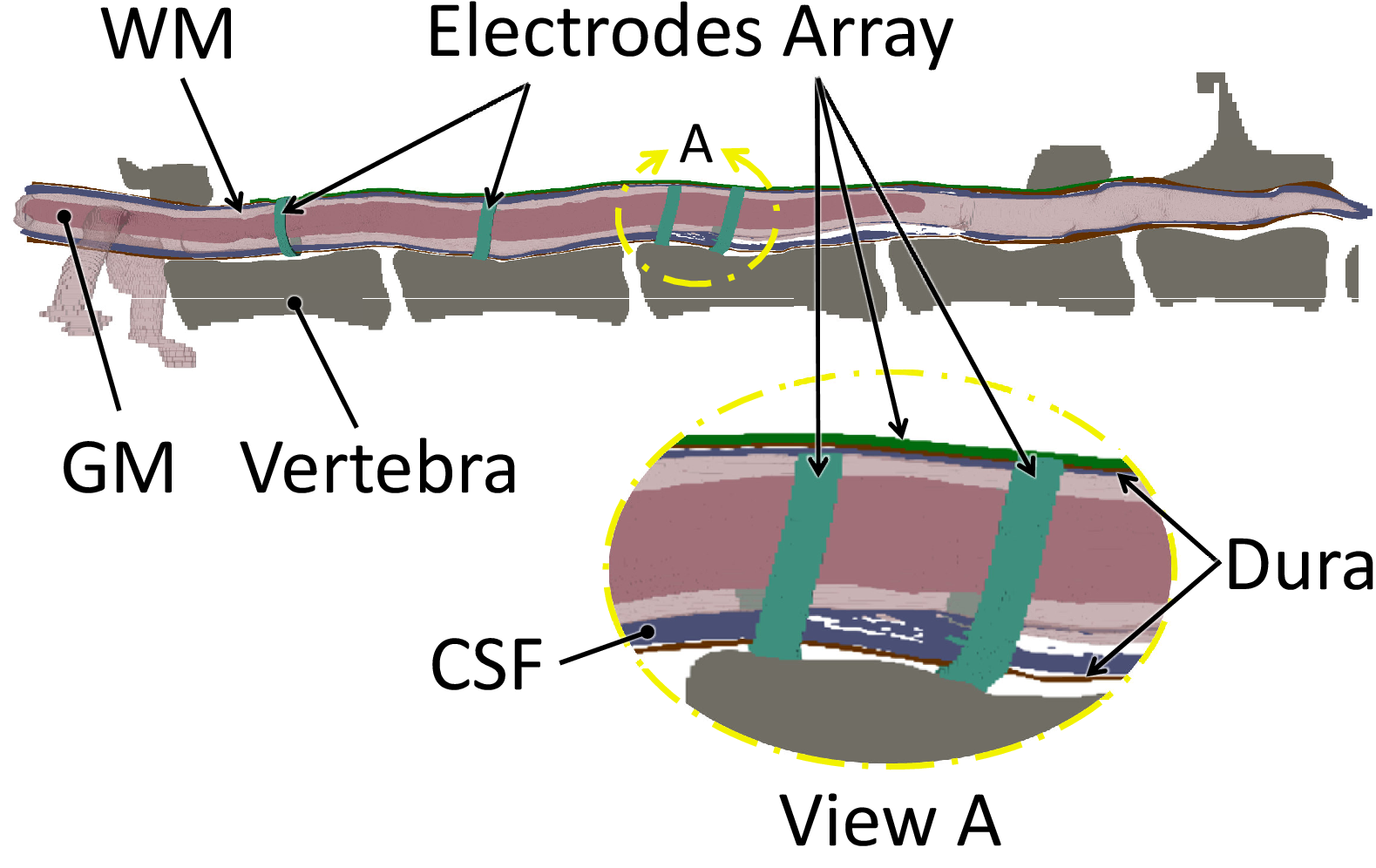}
	     \caption{Cut view of a reconstruction of a cat lumbar spinal cord from CT and MRI images done with paraviewer as to show the different tissues making a spinal cord and electrode placement after surgery. The dura is brown (view A), the CSF is blue (View A), WM is light pink while GM is dark pink. The green rectangle represents the dorsal electrodes pad and the turquoise belts are the ventral electrodes pads. Human geometry is used in calculation. The human spinal cord dimensions used are extracted from \cite{Hong2011,Papinutto2015,Park2016,Calabres2018,Kinaci2020,Toossi2021} and presented supplementary material \cite{SM26T} }
		\label{FIG1}
\end{figure}

	Spinal cord stimulation (SCS) is a specific application in which electrodes are typically positioned on the dorsal surface of the spinal cord, within the spinal canal— composed of dura mater (dura), cerebrospinal fluid (CSF), white matter (WM), and grey matter (GM)— constrained circumferentially by the bony vertebrae  (figure \ref{FIG1}) \cite{Bossetti2008,Labrado2011,Capogrosso2013}. SCS is used routinely for the treatment of chronic pain and is increasingly being investigated for the restoration of movement after paralysis. Therefore, optimizing this therapy requires an understanding of how stimulation affects the target neural structures and how  the EF propagates in and around the spinal cord.
	Mathematical descriptions of the neuronal potential activation threshold with the application of an EF date at least as far back as 1986 \cite{Rattay1986} and form the basis of contemporary models treating how an EF stimulates different tissues \cite{McIntyre2000,McIntyre2002,Capogrosso2018,Patrick2024,Rogers2024}. These results describing spinal cord activation during SCS \cite{Bossetti2008,Arle2013} led to semi-empirical and semi-hybrid computational models that discuss fiber activation by a propagating applied field \cite{Capogrosso2013}. In turn, this inspired computational modeling \cite{Zander2020}, making the correlation between patient response and experimental stimuli possible.

	Together, these modeling approaches were successfully used to predict spinal cord stimulation targets for movement generation in paralyzed individuals \cite{Capogrosso2016,Capogrosso2018,Wagner2018,Rowald2022}. Although these studies have considerably improved empirical tools for spinal cord stimulation modeling \cite{Capogrosso2016,Capogrosso2018,Greiner2020,Solanes2021}, they remain largely based on experimental evidence and results. This limits our ability to probe the fundamental mechanisms of action of spinal cord stimulation, which remain debated \cite{Jensen2018}.
	
	The purpose of this paper is to use analytical continuous calculation of electric field transmission to explore the normal incident EF transmission of a stimulation performed in the vicinity of the spinal cord and to discuss its behavior and propagation. Using electromagnetic concepts, we first describe key physical characteristics of biological tissues, such as complex permittivity  $\tilde{\varepsilon}$, complex wave vector $\tilde{k}$, and complex refractive index $\tilde{n}$, then derive the transmission coefficient $T$. These quantities ($\tilde{\varepsilon}$, $\tilde{k}$, and $\tilde{n}$)  are then used in the equation of transmission $T$ of an applied field through the spinal canal. Our hypotheses are that 1) weak transmission of the EF will be obtained after passing through three layers of tissue; 2) medium transmission of the EF will be observed after a passage through two layers; and 3) the CSF will absorb most of the EF energy. In this context, we propose a theoretical approach to describe and explore the transmission coefficient $T$ mechanism of an externally applied EF by a pair of electrodes.
	
	\section{Methods}
	
	
	\subsection{Physical properties}
	
\begin{figure}[h!t]
	    \centering
	    \includegraphics[ width=8.5cm]{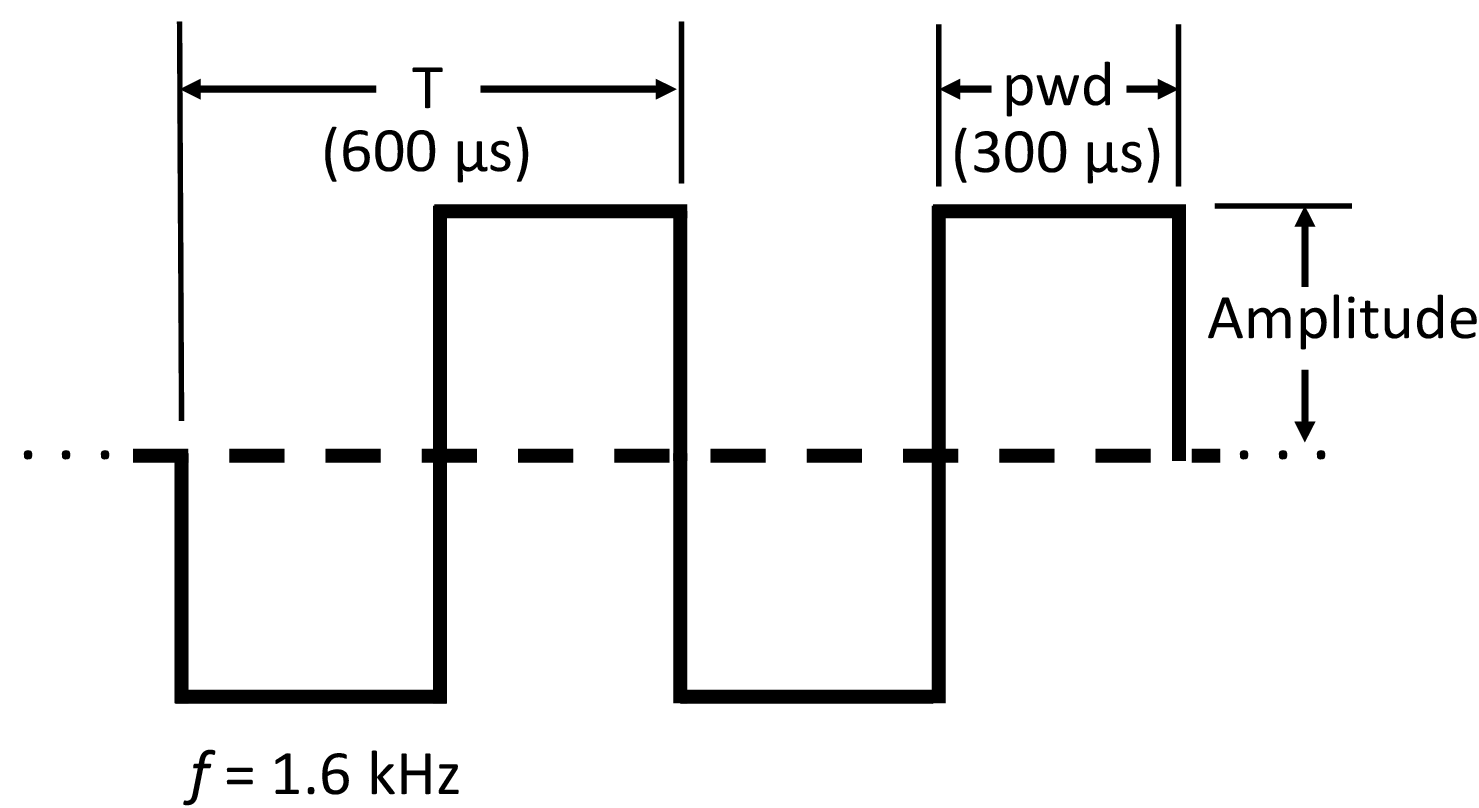}
	     \caption{Schematization of an ideal square wave with $f = 1.6 $ kHz. 
	     The dashed line is the oscillation axis. Here,  pwd is the pulse width.}
		\label{FIG2}
	\end{figure}

In this paper, the theoretical discussion will be based on the lumbar spinal cord of a cat implanted with dorsal and ventral epidural electrode arrays allowing circumferential stimulation and recording. The anatomy is reconstructed from high-resolution microCT images and MRI acquisitions (figure \ref{FIG1}). In addition, we describe pulses as continuous waves as illustrated in figure \ref{FIG2}. We are using the frequency $f$ (in Hz), which is the number of times a pulse repeats itself per cycle within a second. 
Here, the pulse width (pwd) is half the period T of a cycle, which we can relate to the frequency $f$ with the relation $f = 1/(2$ pwd$)$. The wavelength $\lambda$ is the length of the pulse in meters (m) and is $\lambda = v /f$, with $v$ the velocity (m/s) of the pulse in the medium. Permittivity, although usually considered a real number, can be expressed as a complex quantity ($\tilde{\varepsilon}_r$) with a real ($\varepsilon'$) and a complex ($\varepsilon''$) part. The paper will use complex representation.

As neurons and axons can be described as chemical capacitors \cite{Coburn1985,Rattay1986,Capogrosso2013,Naess2020,Rogers2022}, the behavior of the bulk tissue can be approximated as polar dielectrics, which are described by electromagnetic theories \cite{Marion1995,Jackson1999,Borne2002,Sadiku2007,Griffith2013}. 
The spinal cord and surrounding regions can be described as layers of complex dielectric mediums responding to electromagnetic stimuli.

	\subsection{Tissue permittivity}
	
\begin{figure}[h!t]
	    \centering
	    \includegraphics[ width=8.5cm]{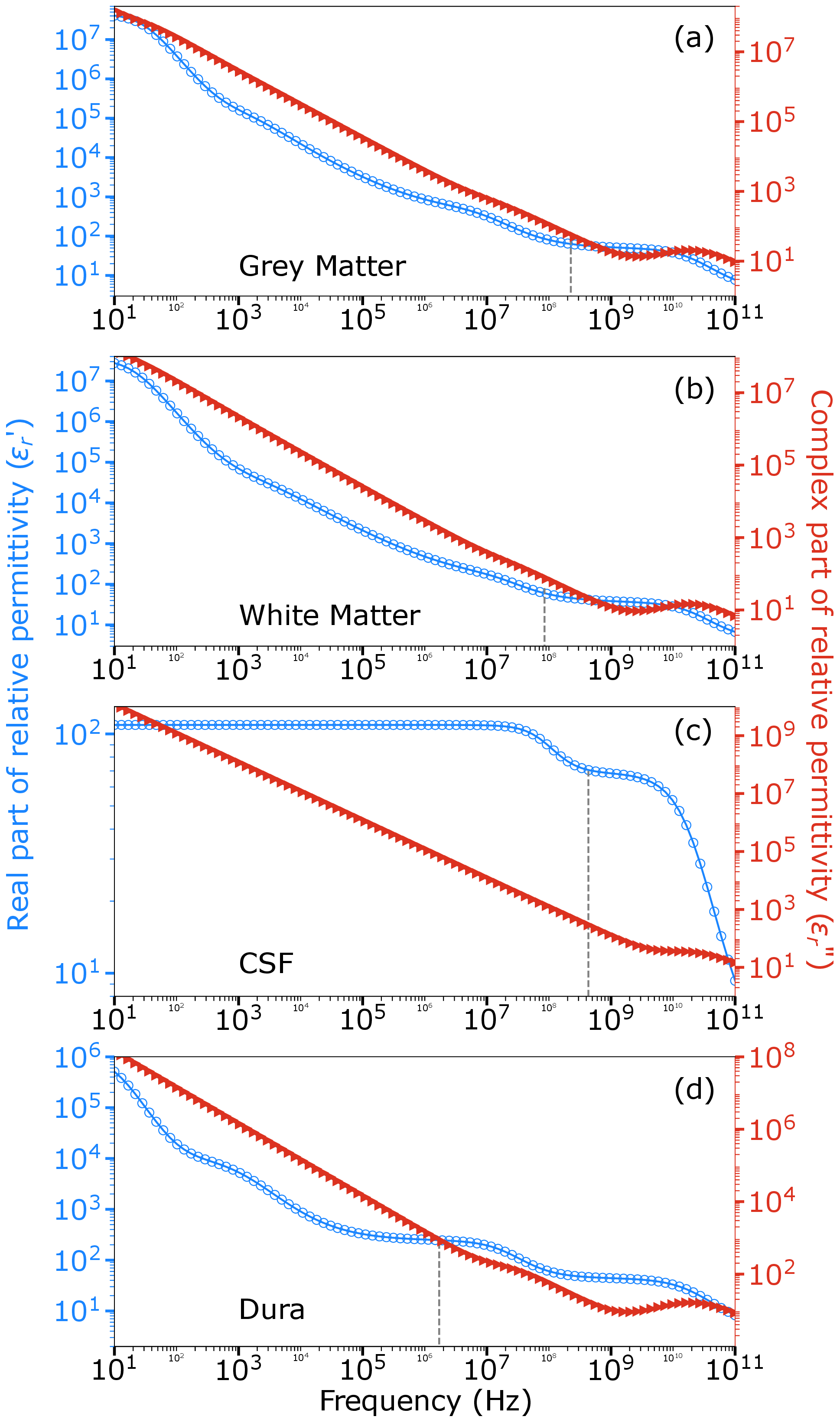}
	     \caption{Calculated real (blue circles- left y axes) and complex (red triangles- right y axes) part of permittivity of (a) Grey matter, (b) White matter, (c) CSF, and (d) Dura tissues. The vertical grey lines indicate where the real and complex parts are equal.}
		\label{FIG3}
	\end{figure}

	The first property to describe for such medium is the relative permittivity $\tilde{\varepsilon}_r$ which is complex due to the lossy nature of biological tissues. The work of S. Gabriel \cite{GabrielS1996a,GabrielS1996b} and C. Gabriel \cite{GabrielC1996,GabrielC2009} on dielectric properties of human tissue provides rich information on $\tilde{\varepsilon}_r$ and conductivity $\sigma$, in addition to being a good starting point. Their work relies on the modified Cole-Cole equation \cite{Pethig1987,GabrielC1996,GabrielS1996a,GabrielS1996b,GabrielC2009}: 
	\begin{eqnarray}
		\tilde{\varepsilon}_r (\omega)= \varepsilon_\infty + \sum\limits_n \frac{\Delta \varepsilon_n}{1 +(\rmi \omega \tau_n)^{(1-\alpha_n)}} + \frac{\sigma_n }{\rmi \omega \varepsilon_0 },
		\label{eq:PErmittivityCole}
	\end{eqnarray}
	\noindent with $\rmi$ the complex number ($\rmi^2 =-1$), $\tau_n$ a time constant characterizing a polarization mechanism of a relaxation region of the dielectric spectrum\cite{Foster1989}. Then, $\varepsilon_\infty$ is the permittivity at $\omega  \tau >> 1$, $\varepsilon_s$ is the permittivity at $\omega \tau <<1$, $\Delta \varepsilon_n = \varepsilon_\infty - \varepsilon_s$ , $\alpha_n$ is the measure of the broadening of a dielectric regime region, $\sigma_n$ is the static ionic conductivity and $\varepsilon_0$ is the permittivity of free space \cite{GabrielC1996,GabrielS1996a,GabrielS1996b}. The calculated values of $\tilde{\varepsilon}_r$ real and complex parts for the studied tissues are presented in figure \ref{FIG3}. The complex part of $\tilde{\varepsilon}_r$ is yet to be reported exhaustively in the literature and little work is done on $\tilde{\varepsilon}_r$ in the frequency region of 10 Hz to 0.1 THz.
	
	Electromagnetism theories state that the greater the complex part, the greater the loss (or absorption), hence the weakening of a propagating pulse, much like in a lossy medium such as a bad conductor or a bad dielectric. The observed results are in agreement with the findings of previous work\cite{Foster1989,GabrielS1996b,GabrielS1996a,GabrielC1996,Peynman2009,Nowak2011,Sollehudin2014,Gun2017,Cherkasova2021,Chen2021,Matkovic2022,Sasaki2022,Kordic2023}. 
	
	The parameters reported by Gabriel \textit{et al.} (1996) were used and extended by fitting the calculations on data from preceding work, allowing us to expand the frequency range below 400 MHz, which is yet to be reported.
	
\subsection{Quasi-static model}
\label{secqasis}
	With the permittivity $\tilde{\varepsilon}_r$ of the tissues well defined, we can look at how an EF propagates within the spinal cord and surrounding tissues. At low frequencies, the wavelength $\lambda$ is multiple order of magnitude the size of the medium, and ohmic boundaries are dominant. In these conditions, traveling waves and refractive boundary reflections do not physically manifest. Hence, we must use the quasi-static description of the EF to describe its influence on-- and propagation within-- said medium. In this regime, one can use Laplace volume conduction \cite{Plonsey1967,Hamalainen1993,Taulu2021,Caussade2023}:
\begin{eqnarray}
	\nabla^2 \phi = 0 ,
\end{eqnarray}
that has a solution of the following form:
\begin{eqnarray}
	\phi = A x + B .
\end{eqnarray}
Using this solution, potential conservation and current density conservation, we can write \cite{Heubrandtner2004}:
\begin{gather}
	\begin{bmatrix}
		A _i \\
		B_i
	\end{bmatrix}
	=
	\begin{bmatrix}
	\sigma_j / \sigma _i & 0 \\
	r_m (1 - \sigma_j/\sigma_i ) & 1
	\end{bmatrix}
	\begin{bmatrix}
		A_j \\
		B_j 
	\end{bmatrix} .
\end{gather}
Generalizing for three junctions (four tissues), we get:
\begin{gather}
	\begin{bmatrix}
		A_1 \\
		B_1
	\end{bmatrix}
	=
	C_1 C_2 C_3
	\begin{bmatrix}
		A_4 \\
		B_4
	\end{bmatrix} .
\end{gather}

Solving this matrix multiplication, using the fact that $E_{initial} = V_0 / r$ and $T = |E_{target} / E_{initial}|$, one can get :
\begin{eqnarray}
	 T_{QS}  = \left| \frac{r_1 }{ r_1 M_{11} + M_{21} } \right|^2 ,
	 \label{eq:QS6}
\end{eqnarray}
with $QS$ standing for $quasi-static$. Here, $M_{11}$ stands for
\begin{eqnarray}
	M_{11} = \frac{\sigma_4}{\sigma_1 } ,
\end{eqnarray}
and $M_{21}$ for
\begin{gather}
	M_{21} = r_1 \left( \frac{\sigma_4}{\sigma_2} - \frac{\sigma_4}{\sigma_1} \right) + r_2 \left(\frac{\sigma_4}{\sigma_3 } - \frac{\sigma_4}{\sigma_2} \right) + r_3 \left( 1 - \frac{\sigma_4}{\sigma_3} \right) .
\end{gather}
The detailed derivation of these relations is presented in supplementary material \cite{SMQS26}. Since spinal cord tissues physical properties are frequency sensitive, we use their frequency dependent value of conductivity in the calculation of $T_{QS}$ \cite{GabrielC2009,Caussade2023}. Finally, for the transmission between two layers, the quasi-static Fresnel equation is used \cite{Skaar2019} :
\begin{gather}
	T_{QS \text{- 2 layers}} = \frac{ k_{2z} }{ k_{1z} }  \left| \frac{ 2  k_{1z} }{ k_{1z} + k_{2z} } \right|^2 ,
	\label{eq:QS2c}
\end{gather}
with the $k_{iz}$ the ones described in the cited article.

	\subsection{Complex wave vector}
	
	 With the quasi-static model defined, we can explore the higher frequencies range where the spinal cord size is not neglegible in front of $\lambda$. In this regime, we can use the wave description of the EF in matter. In a given medium, a propagating EF can be described using sinusoidal solution\footnote{A square pulse would be described as an infinite sum of sinusoidal functions. Here we will describe what happens to all the components of said square pulse by calculating the transmission coefficient of each pulse frequency composing it.} such as \cite{Jackson1999,Griffith2013,Khalid2015}:
	\begin{eqnarray}
	 \mathbf{E} = E_0\rme^{\rmi (\tilde{k} \mathbf{r} - (2 \pi f) t) } \uvec{e}_i ,
	\end{eqnarray}
	with the complex wave vector $\tilde{k} = k +i \kappa$, which are the phase velocity and extinction coefficient, respectively. 
	Both parts can be described as \cite{Boylestad1985,Marion1995,Jackson1999,Borne2002,Griffith2013,Sadiku2007}:
	 \begin{eqnarray}
	 	k (f) = \frac{2 \pi f }{c} n'_i (f ), \\
	 	\kappa (f) = \frac{2 \pi f }{c} n''_i (f),
	\end{eqnarray}
	with $\tilde{n}_i$ the complex refractive index for the $i$\textsuperscript{th} tissue, $n_i'$ and $n_i''$ being the real and imaginary part of the index, respectively, and $c$ the speed of light in vacuum.
	
	\subsection{Complex refractive index}
	
	The refractive index is proportional to the permittivity and magnetic permeability as\cite{Jackson1999,Griffith2013,Cherkasova2021}:
	\begin{eqnarray}
		\tilde{n} (f) = \sqrt{ \frac{ \tilde{\varepsilon}  (f ) \tilde{\mu}( f )}{\varepsilon_0 \mu_0} }  .
	\end{eqnarray}

	One can write relative magnetic permeability ($\mu_r$) proportional to magnetic susceptibility ($\chi_{M}$) as $\mu_r = (1+ \chi_{M})$ \cite{Marion1995,Jackson1999,Borne2002,Sadiku2007,Griffith2013}. In addition, previous work on the $\chi_{M}$ of biological tissue \cite{Vorauer2016,Klohs2021} shows that the susceptibility of human tissues is of the order of magnitude $\chi_M \sim 10^{-8}$ m$^2$ kg$^{-1}$, which is small enough to write $\mu\simeq \mu_0$ even at low frequencies. This is due to the fact that the EF intensity of interest/used in clinical situations will not be sufficiently strong to generate a great enough magnetic field to see a significant magnetic response. This leads us to:
	\begin{eqnarray}
	        \tilde{n} (f)= \sqrt{ \tilde{\varepsilon}_r  (f) }.
	        \label{eq:netvarepcompsqrt}
	\end{eqnarray}
	Using (\ref{eq:netvarepcompsqrt}) and the fact that permittivity ($\tilde{\varepsilon}_r = \varepsilon'_r + i \varepsilon''_r$) as well as the refractive index ($\tilde{n} = n' + i n''$) are complex, we can use $n'$, $n''$, $\varepsilon'$, and $\varepsilon''$ as in the Appendix 
	to get the following relations:
	\begin{eqnarray}
		n' (f ) = \left\lbrack \frac{1}{2}\left(  (\varepsilon_r'^{ 2} +\varepsilon_r''^{  2})^{1/2} +  \varepsilon_r' \right) \right\rbrack^{\zeta' /2}, \label{eq:refrar}\\
		n'' (f ) = \left\lbrack \frac{1}{2}\left(  (\varepsilon_r'^{ 2} +\varepsilon_r''^{  2})^{1/2} -  \varepsilon_r' \right) \right\rbrack^{\zeta'' /2}, \label{eq:refraim}
	\end{eqnarray}

	with variables $\zeta'$ and $\zeta''$ present to fit calculations to the scarcity of data from previous work \cite{Penfold1929,Jacques2013,Zhu2018,Nagel2018}, 
	
\begin{table}[ht]
\caption{\label{table:Param1}%
Refractive index variable values for the different tissues.}
\begin{ruledtabular}
\begin{tabular}{lccdr}
\textrm{$\zeta$}&
\textrm{GM}&
\textrm{WM}&
\textrm{CSF}&
\textrm{Dura}\\
\colrule
	$\zeta'$ & 0.435 & 0.5 &  0.42 & 0.55 \\ 
	$\zeta''$ & 2 & 2 & 1 & N/A \\
\end{tabular}
\end{ruledtabular}
\end{table}	
\noindent with the exception of the complex part of dura, which behaves as an exponential with an inflection point near 60 THz, which justifies the multiplication of its $n''(f)$ by $ \text{exp}\lbrack - f \cdot 63.4 \times 10^{-12} \rbrack$. The variables $\zeta'$ and $\zeta''$ are exponents, in a similar fashion to $\alpha_n$ of the modified

\noindent    Cole-Cole equation \cite{Pethig1987,GabrielS1996b,GabrielS1996a,GabrielC1996}. This variable is introduced for $\tilde{n}$ rather than $\tilde{\varepsilon}_r$ as it is already well defined by the modified Cole-Cole equation. The values of $\zeta'$ and $\zeta''$ are presented in table \ref{table:Param1}. These equations are valid for a propagating electromagnetic field in the high frequencies region, as the magnetic properties of matter do not have time to react. They would not hold at lower frequencies if $\chi_M$ was great, as the magnetic interaction would have enough time to influence the system. However, as mentioned earlier, it is reported that human tissues $\chi_M$ are too weak for the used EF to generate a magnetic response. Hence, we assume that the previous equations are valid at lower frequencies.

\begin{figure}[ht!]
	    \centering
	    \includegraphics[ width=8.5cm]{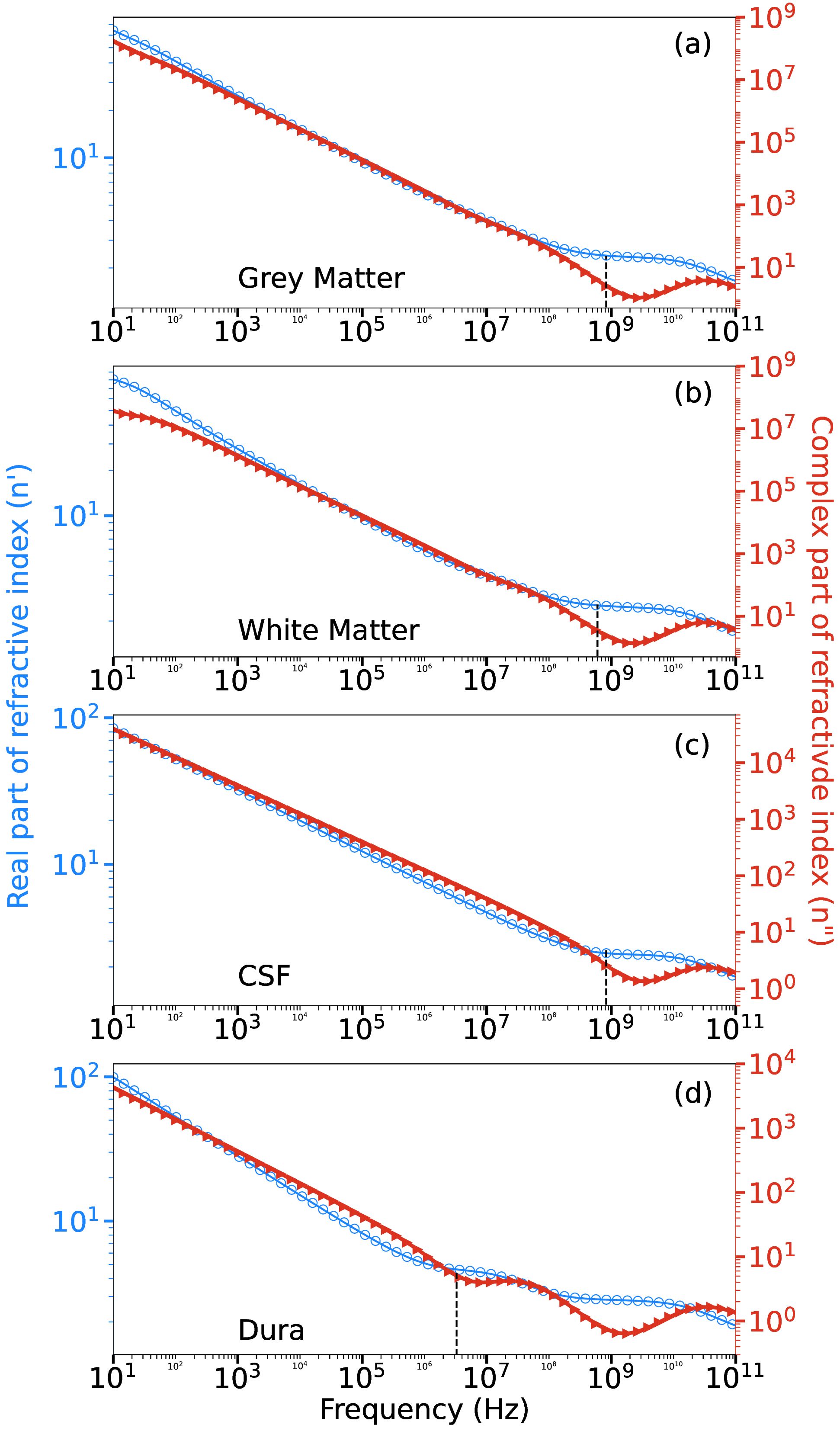}
	    \caption{Calculated real (blue circles- left y axes) and complex (red triangles- right y axes) part of refractive index of (a) Grey matter, (b) White matter, (c) CSF, and (d) dura tissues. Vertical grey lines indicate the frequency at which the real and complex parts are equal.}
		\label{FIG4}
\end{figure}

	 The results of the calculations of $n'$ and $n''$ of the different tissues of the spinal cord are shown in figure \ref{FIG4}, which has yet to be extensively reported for frequencies below 0.1 THz. The calculated values at higher frequencies are in agreement with previous work \cite{Penfold1929,Yaroslavsky2002,Gebhart2006,Jacques2013,Zhu2018,Nagel2018,Wang2020,Shapey2021}. More measurements at low $f$ are required to confirm these results. 

Even so, values at higher frequencies are slightly higher than reported  values\cite{Penfold1929,Jacques2013,Wang2020,Shapey2021,Yaroslavsky2002,Gebhart2006,Zhu2018,Nagel2018}. Still, the theoretical values presented are valid as they are for human tissues and not other animal tissues (such as bovine, pig, rat, or mouse as examples) like the ones reported in the literature, which often differ from humans. It is good to underline the need for a greater quantity of data at lower frequencies to confirm the calculated results, down to the low radio frequencies (lRF).

	\subsection{Impedance}
	
	Finally, even if at first we are interested in the transmission coefficient $T$, there is reflection of the EF within the system due to characteristic impedance mismatch $\eta$ between the different tissues. Such an impedance can be written as:
	\begin{eqnarray}
		\eta = \frac{\tilde{\mu}}{\tilde{n}},
	\end{eqnarray}
		as discussed earlier, we can assume $\tilde{\mu}\rightarrow \mu_0$ in the given tissues and frequency range explored. In addition, we take $\mu_0$ as unity. We then get:
	\begin{eqnarray}
		\eta (f) \simeq \frac{1}{\tilde{n} (f)} .
		\label{eq:impedance}
	\end{eqnarray}
	Hence, the higher the refractive index, the lower the characteristic impedance. If the impedance of the first medium is greater than that of the second ($\eta_1 > \eta_2$), the reflection of the EF is hard and is reflected with a phase difference of a factor $\pi$ with the incident field, resulting in a lower proportion of the EF being transmitted. In the opposite case ($\eta_1 < \eta_2$), the reflection of the EF will be soft and the reflected field will be in phase with the incident field.

\subsection{Wave propagation model}
\label{sec:waveprop}

	Since most SCS modeling studies reported in the known literature use the finite element method \cite{Capogrosso2013,Jensen2018,Patrick2024}, we would like to complement existing tools with an updated theoretical model of normal incident EF propagation in the spinal cord and surrounding regions at higher frequencies ($> 1\freq{M}$). This model, applied exclusively as a radiation absorption model in the high-frequency regime, will address the lack of a more theoretical description of EF propagation, considering real and complex components of the system elements, through spinal cord tissues. This could help us better understand the spinal cord's reaction to an applied EF and open avenues for different approaches to SCS.

\begin{figure}[h!t]
	    \centering
	     \includegraphics[ width=8.5cm]{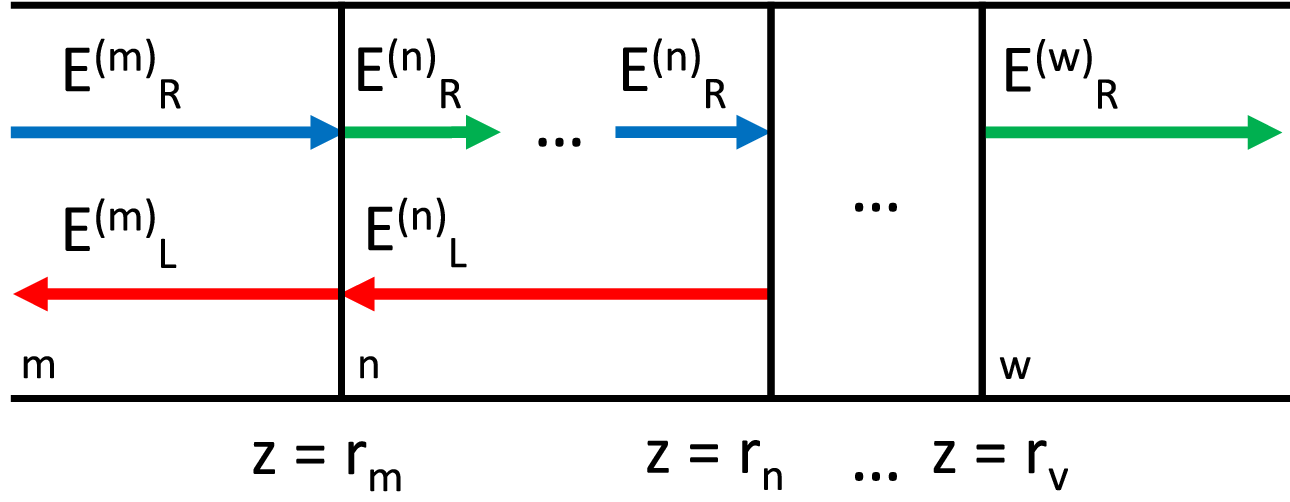}
	     \caption{Schematization of the electric field direction of propagation and position in a medium with multiple junctions. The vertical black lines represent the different junction.}
		\label{FIG5}
\end{figure}

	The SCS paradigm most frequently used in clinical practice consists of a monophasic or biphasic square pulse current (at $f \sim 1.6 \freq{k}$) injected by a pair of electrodes\footnote{It is good to mention that epidural stimulation can be done at higher $f$, up to 20 kHz.}. This generates an EF in the epidural space \cite{Bossetti2008,Capogrosso2013,Lempka2015,Zander2020,Rowald2022,Rogers2024} that will propagate through the tissues in the vicinity of the spinal cord. These include dura, CSF, WM, and GM. Because the physiological target of stimulation are the neurons of the spinal cord, it is important to know how much of the applied EF reaches the WM and GM. Using the full complex wave vector $\tilde{k}$, the boundary conditions can be written as sinusoidal solutions
\begin{gather}
		 E_\RR^{(m)} \rme^{\rmi k_m r_m -\kappa_m r_m} +E_\LL^{(m)}\rme^{-\rmi k_m r_m +\kappa_m r_m} 
		 = E_\RR^{(n)}\rme^{\rmi k_n r_m-\kappa_n r_m} +E_\LL^{(n)}\rme^{-\rmi k_n r_m +\kappa_n r_m} ,
	\end{gather}
	\begin{gather}
	E_\RR^{(m)}\rme^{\rmi k_m r_m -\kappa_m r_m} - E_\LL^{(m)} \rme^{-\rmi k_m r_m +\kappa_m r_m}  
	    = \frac{\tilde{n}_n}{\tilde{n}_m}(E_\RR^{(n)}\rme^{\rmi k_n r_m -\kappa_n r_m} -E_\LL^{(n)} \rme^{-\rmi k_n r_m +\kappa_n r_m}),
\end{gather}
	with subscripts $R$ (Right) and $L$ (Left) giving the propagation direction of the wave, and $r_{\text{m}}$ the position in space, as shown in figure \ref{FIG5}. 
	
	The time-dependent argument of the exponentials cancels out at the boundaries as they are at the same $t$ and $f$. Using those, considering continuity, and specifying that the incident pulse is normal to the plane of incidence, we can write the matrices for each interface of two mediums such as:
	\begin{gather}
	 \underbrace{
	    \begin{bmatrix}
	        \rme^{\rmi k_{m} r_{m}-\kappa_m r_m }& \rme^{-\rmi k_m r_m +\kappa_m r_m } \\
	        \rme^{\rmi k_{m} r_{m}-\kappa_m r_m } & -\rme^{-\rmi k_m r_m +\kappa_m r_m }
	    \end{bmatrix}}_{M_j}
	    \begin{bmatrix}
	        E_\RR^{(m)} \\ E_\LL^{(m)}
	    \end{bmatrix} 
	  = \underbrace{\begin{bmatrix}
	       \rm \rme^{\rmi k_{n} r_{m} -\kappa_n r_m} & \rme^{-\rmi k_{n} r_{m} +\kappa_n r_m} \\
	        \frac{\tilde{n}_n}{\tilde{n}_m}\rme^{\rmi k_{n} r_{m} -\kappa_n r_m} & \frac{-\tilde{n}_n}{\tilde{n}_m}  \rme^{-\rmi k_{n} r_{m} +\kappa_n r_m }
	    \end{bmatrix}}_{N_j}
	    \begin{bmatrix}
	        E_\RR^{(n)} \\ E_\LL^{(n)}
	    \end{bmatrix} ,
	\end{gather}
	which can be written as:
	\begin{eqnarray}
	    M_j E_i = N_j E_{i+1} .
	\end{eqnarray}
	    Repeating the exercise for each interface, we can get the scattering matrix \textbf{F} \cite{Balili2012} for four layers:
	\begin{eqnarray}
	    E_1 =  M_1^{-1} N_1 M_{2}^{-1} N_{2} M_{3}^{-1} N_{3} E_{4} = \mathbf{F} E_{4}.
	\end{eqnarray}
	Completing the matrix product, we get the detailed transfer matrix $\mathbf{F}$. Then, we can use the relation:
	\begin{eqnarray}
		T_{wp} = \frac{I_{\text{T}}  }{I_0 } = \frac{\tilde{n}_{\text{T}} }{\tilde{n}_0} \left| \frac{E_T }{ E_0} \right|^2 ,
	\end{eqnarray}
	where $wp$ stands for wave propagation. We use this relation to get
	\begin{eqnarray}
		T_{wp} =  \frac{\tilde{n}_4}{\tilde{n}_1}  \frac{4}{ (2 \text{Re}\lbrace F_{00} \rbrace)^2 +(2 \text{Im}\lbrace F_{00} \rbrace)^2  },
		\label{eq:transmissiondansTheory}
	\end{eqnarray}
	which depends on $\tilde{n}$, $\tilde{k}$, $f$, and tissue thickness. The detailed expression of $F_{00}$ is written in the supplemental material \cite{SM26}.
	
	Finally, one should describe the propagation of EF when there are only three layers of tissue as well. This will allow us to know how much of EF energy reaches the WM and how much reaches the GM when it touches the CSF at the dorsal horn of the spinal cord. The same method as for four layers is used, stopping the calculation at two junctions instead of three as previously. This is written as another transfer matrix named $\mathbf{F}'$, and leads to \footnote{It is important to underline that only the real part of $\tilde{n}_j / \tilde{n}_i$ are used in equations \ref{eq:transmissiondansTheory} and \ref{eq:transmissiondansTheory3c} as $T$ is a real quantity.}:
	\begin{eqnarray}
		T_{wp} =  \frac{\tilde{n}_3}{\tilde{n}_1}  \frac{4}{(2 \text{Re}\lbrace F'_{00} \rbrace )^2 +(2 \text{Im}\lbrace F'_{00} \rbrace )^2} ,
		\label{eq:transmissiondansTheory3c}
	\end{eqnarray}
	which is of the same form as the previous equations. The procedure of calculation to get $F_{00}'$ is written in the in the supplemental material \cite{SM26}.  

	For $T_{wp}$ from dura to CSF, the Fresnel equation for two tissues is used, which we multiply by Beers law to take into account absorption by the dura as (here, D stands for dura):
	\begin{eqnarray}
		T_{wp \text{- 2 layers}} = \frac{ \tilde{n}_{\text{CSF}} }{ \tilde{n}_{\text{D}} }\left| \frac{2\tilde{n}_{\text{D}}  }{( \tilde{n}_{\text{CSF}} + \tilde{n}_{\text{D}} )} \right|^2 e^{- 2\pi f n_{\text{D} }'' r_m /c }
		\label{eq:deuxtissuT}
	\end{eqnarray}
	
 The models presented in section \ref{secqasis} (quasi-static) and \ref{sec:waveprop} (wave propagation) are then run using a Python script\cite{python2026} that calculates the value of $T$, using the proper model ($QS$ or $wp$) for a given frequency range, with $f$ going from $10 \freq{}$ up to $0.1 \freq{T}$. 
	
	\subsection{Physiological considerations}
	
	As water is a major element of biological tissues, it is relevant to discuss its physics here. With a very large dipolar moment, in addition to having hydrogen bonds to the ions in the medium, we know that the water content has a significant contribution to the dielectric behaviors and the dipolar moment of the molecular structures \cite{Ma2024}.
	
	 In the frequency range of $10 \freq{}$ to $4 \freq{G}$, we expect water to have little damping effect on tissue vibration modes. In fact, even if the water molecules can rotate freely, they are unified into molecular structures of tissues, moving as a unit with the structure instead of individuals. Such massive structures can only present bulk modes of vibration in the low band of microwaves \cite{Prohofsky2004,Yada2008,Sheppard2008,Sun2021,Mancini2022}. 
	 With an applied EF around $0.1 \freq{G}$, rotation of water molecules starts to be inhibited by their rotative inertia (the ``dielectric friction''). In other words, the absorbed energy is lost via collisions or interactions with the next neighbors (we can also talk of molecular viscosity). Near $f\sim 4 \freq{G}$, the water starts to exhibit a larger damping of the molecular modes, and there is an increase in the viscosity of the medium. At 10 GHz, water becomes overdamped and inhibits the vibrational modes of molecules. From then on, the water damping effect will weaken with frequency, as can be seen by its dielectric constant at infinity ($\varepsilon_{\infty}=1.8$) that tends toward a frequency-independent behavior \cite{Yada2008,Sheppard2008}. Then, in the hundreds of GHz ranges, from 300 GHz and higher, we observe molecular resonance and vibration modes that can absorb EF energy.
	
	We expect all of this to be truer for tissues with a greater part of water constituting them, such as CSF. In tissues with a lower water content, we still expect a strong influence on their behavior by the water element, but with a greater proportion of absorption linked to the presence of proteins, lipids, and other constituents present.
	
\section{Results}
	
\begin{figure}[ht!]
	\begin{subfigure}[b]{16cm}
		\centering
	     \includegraphics[ width=8.6cm]{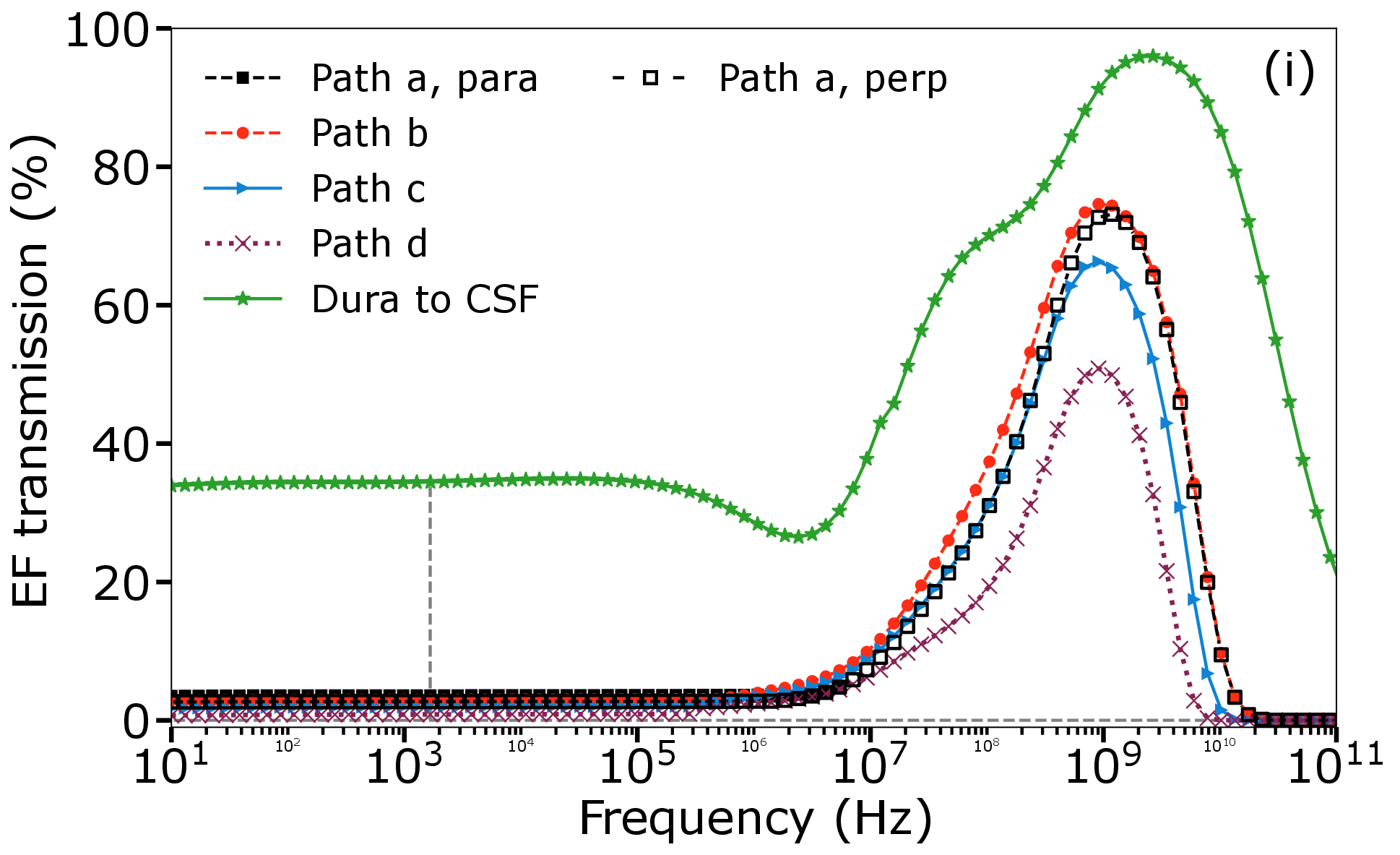}
	\end{subfigure}
	\begin{subfigure}[b]{4.25cm}
	     \centering
	        \includegraphics[width=4.25cm]{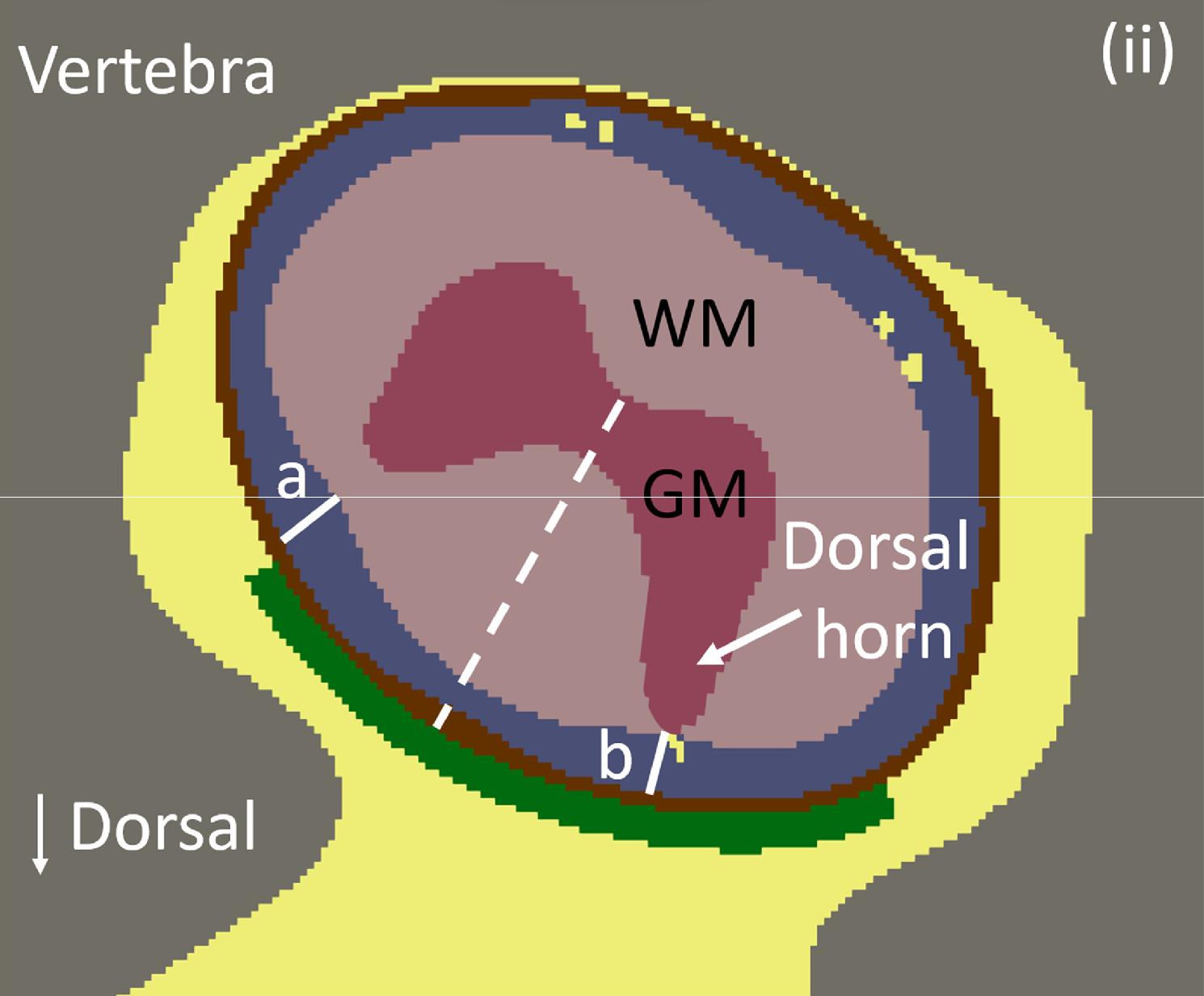}
	    \end{subfigure} 
	\begin{subfigure}[b]{4.25cm}
	     \centering
	        \includegraphics[width=4.25cm]{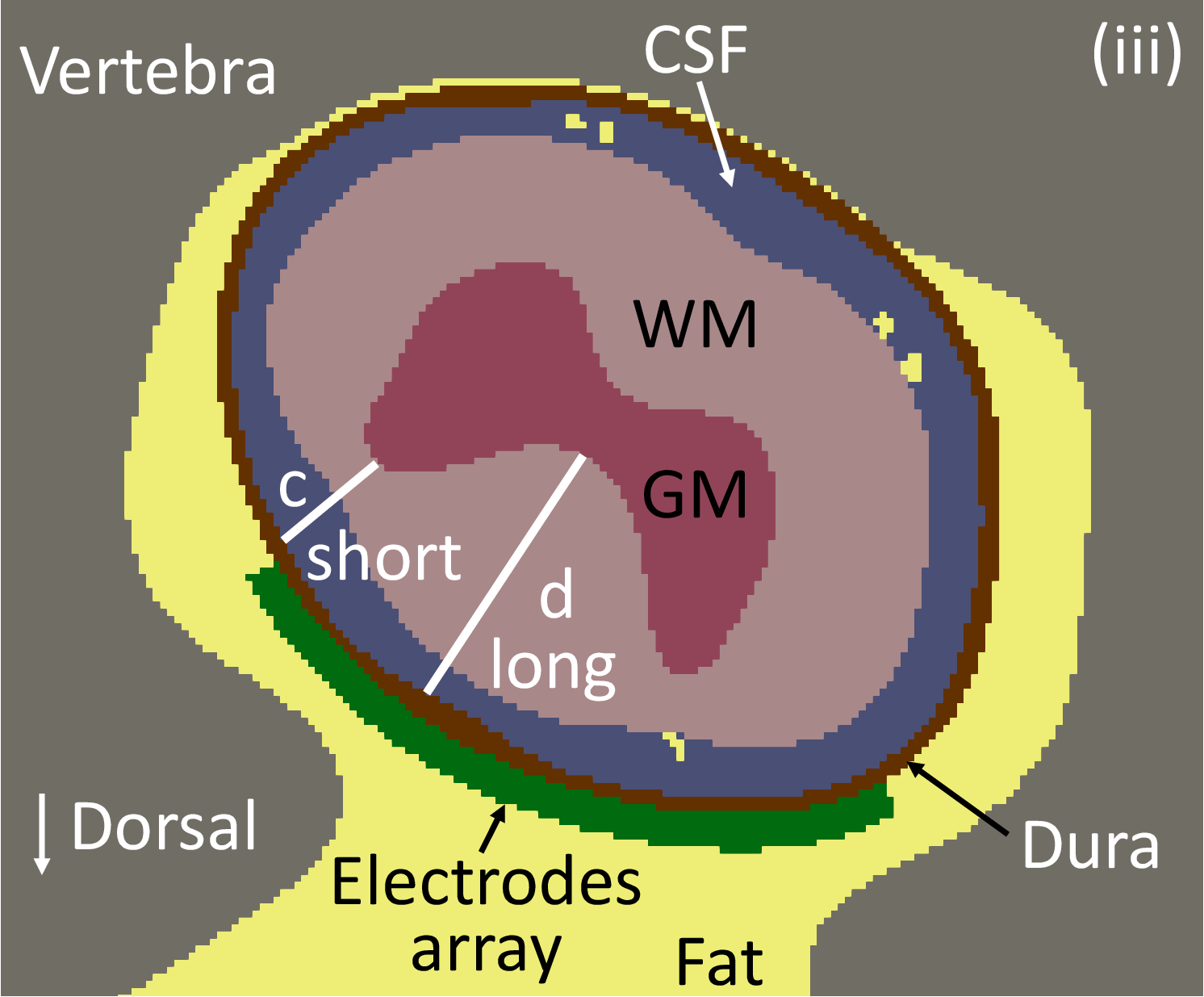}
	\end{subfigure} 	   
	     \caption{ (i) Calculated transmission of the EF to the last tissue of different path according to the frequency. The vertical grey line corresponds to the clinically used biphasic pulse centered at 1.6 kHz (pulses up to $30 \freq{k}$ can be used in stimulation) and the horizontal one is a guide for the eye for the origin of the y axis. (ii) and (iii) are slice views of a cat reconstructed spinal cord with the electrodes (green) installed on the dura (brown) tissue. The white lines a (para for parallel to the WM fibers and perp for perpendicular to the fibers) and b are the EF paths through two layers targetting white and grey matter respectively (both $\sim 2.78 \text{ mm}$), whereas c ($\sim 4 \text{ mm}$) and d ($\sim 7.3 \text{ mm}$) are the EF paths going through three layers both targetting grey matter. The dashed line in (ii) is there to separate two possible form of the GM. Human geometry is used for calculation. The human tissues measurements used are extracted from previous work \cite{Hong2011,Papinutto2015,Park2016,Calabres2018,Kinaci2020,Toossi2021} and presented supplementary material \cite{SM26T}.}
	\label{FIG6}
\end{figure}

Using the results of figure \ref{FIG4} as well as of table \ref{table:Param1} in (\ref{eq:QS6}), (\ref{eq:QS2c}), (\ref{eq:transmissiondansTheory}), (\ref{eq:transmissiondansTheory3c}), and (\ref{eq:deuxtissuT}), we calculate the transmission coefficient of a propagating EF with normal incidence and for different paths in tissues (lines a and b are the paths through two layers targetting white and grey matter respectively whereas c and d are the paths through three layers both targetting grey matter) which is presented in figure \ref{FIG6}.   
It is important to underline that these results are theoretical in nature and its validation limited to existing experimental results\cite{GabrielS1996b,GabrielS1996a,GabrielC1996,Penfold1929,Jacques2013,Zhu2018,Nagel2018,Shapey2021}. In addition, in an experimental setting, contemporary stimulation devices usable for SCS can generate a stimuli pulse with a frequency that goes up to $f \sim 30 \freq{k}$ \cite{Bocci2018}, which is far from the maximum explored frequency ($0.1 \freq{T}$) of this work. If a stimulation device capable of generating a pulse with a frequency in the GHz region is developed, results in figure \ref{FIG6}  could be used as a point of reference to guide future research on SCS.

	In perfect dieletrics, an increase in $f$ would lead to a linear decrease in impedance, hence a greater EF wave transmission. This behavior is not seen in the results, as the mediums that make up the spinal cord behave more like a conductor than a dielectrics for $f$ lower than a given threshold for each tissue. This is seen by the weak $T$ at low $f$. This gives place to complex interactions dependent on the $f$ of the EF.

\section{Discussion}
\subsection{Transmission between dura mater and CSF}
As the applied EF reaches the junction of dura and CSF, $\sim 34 \text{\%}$ of its energy is transmitted to the CSF for $f$ going from $10 \freq{}$ up to $0.1 \freq{M}$. The great loss  in energy is attributed to induced ionic current in the CSF, as it still behaves as a conductor in this frequency range. At $0.1 \freq{M}$, it is where dura and CSF $\sigma$ start to diverge, leading to a rise of the shunt in the CSF due to its great conductivity. Then, $T$ continues to lower up to $f \sim 2 \freq{M}$, where its slope rises again. In the following, the slope of $T$ decreases at $f \sim  60$ MHz and rises at $f \sim 0.3 \freq{G}$. The first inflection point (at $2 \freq{M}$) coincides where dura behavior goes from conductor to dielectric and the later two are where impedance differences between the two tissues varies, changing $T$ of the EF. Lastly, $T$ for the dura to CSF interface peaks at $f \sim 2.4 \freq{G}$ with $T\sim 96 \text{ \%}$, then decreases due to absorption by the dura and shunting effects at the interface \cite{Prohofsky2004,Sheppard2008}.
	
	\subsection{Transmission through multiple layers, low frequency regime}
	
	In the low-frequency region ($0.1 \freq{k}$ to $1 \freq{M}$), or the quasi-static regime, $T$ for the path a presents two distinct curves since its target tissue, WM is highly anisotropic ($\sigma_x = \sigma_z = \sigma_{perp}$ and $\sigma_y = \sigma_{para}$). The first one is for when the energy is propagating perpendicular to the WM fibers (path a, perp-- empty black squares) which present a $T$ that goes from $ 2.65 \text{ \%}$ to $2.78 \text{ \%}$ as it presents a weaker conductivity. The second solution is when the induced current goes along the aligned myelinated axon tracts of WM (path a, para-- full black squares), which presents a $T$ that goes from $ 3.5 \text{ \%}$ to $ 3.7 \text{ \%}$. Since the CSF behaves as a conductor at this frequency range, we see induced current within the CSF. Then after most of the current is lost within the CSF due to shunting effect. What little energy that reaches WM follows the path of least resistance, which is along the WM fibers at the surface of the junction between CSF and WM. Considering this, we assume that, for $f < 1\freq{M}$, the solution of path a- para is the one that properly describes the EF propagation within the spinal cord with WM as the target tissue. For this reason, it is this conductivity ($\sigma_{WM-para}$) that is used in the calculation of $T$ for path b, c, and d in this frequency range. The $T$ for b goes from $2.8 \text{ \%}$ to $3 \text{ \%}$, $T$ for c goes from $1.75 \text{ \%}$ to $ \text{2 \%}$, and $T$ for d goes from  $0.7 \text{ \%}$ to $0.9 \text{ \%}$.

The differences between the results of $T$ for path a and b, although within $1 \text{ \%}$ from one another, are due to the fact that the target tissues are different for both path, with WM an anisotropic tissue that differs from the isotropic GM. As for the difference between c and d, it is due to the thickness of the WM that has a small but visible impact on $T$. These results underscore that the EF reaches GM or WM with a weak intensity, yet the transmission is greater from CSF to WM (parallel to fibers) than from CSF to GM due to its greater conductivity ($\sigma_{WM} > \sigma_{GM}$) along the fibers. 
	
	For the clinical frequency ($f \sim 1.6 \freq{k}$), the a and b paths exhibit $T\simeq 3.6\text{ \%}$ and  $T \simeq 2.9 \text{ \%}$ respectively, whereas c and d paths exhibit $T\simeq 1.93 \text{ \%}$ and $T \simeq 0.82 \text{ \%}$ respectively. With these results, we can hypothesize that the stimulation EF centered at $1.6 \freq{k}$ is strongly absorbed in dura and CSF as induced current (shunt effect), which also explains the low values of $T$ calculated. In addition, one must consider that some current is coming from the electrodes as well as it follows the path of the least impedance. The total current then interacts with bodies floating in the CSF at first rather than with the GM and/or WM, as more energy reaches them. These results are consistent with previous work observations \cite{Wesselink1998,Bossetti2008,Capogrosso2013,Arle2013,Ladenbauer2010}. 
	
	
	Lastly, the $QS$ description of EF propagation converges to the $wp$ model for frequencies ranging from $0.5 \freq{M}$ (path c and d) up to $12 \freq{M}$ ($T$ from dura to CSF). This frequency dependence of the model convergence is attributed partly to the frequency-dependent properties of the different tissues composing the spinal cord. Additionally to this, the $wp$ model presented assumes a normal incidence and perfectly perpendicular EF. This in turns implies a neglect of 3D geometrical effects on the propagating EF, which results in an overestimation of $T$ for frequencies below $5 \freq{M}$. These limitations are further discussed at the end of this work.

	\subsection{Transmission through multiple layers, intermediate frequency regime} 
	
	Starting from $f \geq 1.6 \freq{M}$ (where $\lambda \lesssim 6.5 \text{ m}$ in the matter) the wave propagation model can be used to some extent to describe $T$. At $1.6 \freq{M}$, the slope of $T$ increases, which is where the dura behavior to an applied EF transitions from conductor to dielectric. This results in a conductor (electrode)-dielectric (dura)-conductor (CSF) scenario, or a capacitor. The slope of $T$ continues to increase for every configuration until the WM behavior changes to dielectric at $f \sim 85 \freq{M}$. It is also where we see a drop in tissues $\varepsilon_r''$ as well as a plateau of $\varepsilon_r'$ and where molecular viscosity starts to rise. As water begins to exhibit dielectric friction, it is appropriate to discuss the behavior of $T$ for the case of four- and three-layer path of the propagating EF in this frequency range.
	
	In the case of four layers, this creates a periodic dielectric-conductor configuration. This means that the system is akin to a capacitor with CSF and GM as conductors, and with dura and WM as dielectrics. However, this state holds up to $f \sim 0.2 \freq{G}$ where GM behavior changes to the one of a dielectric, which makes the capacitor description of the situation insufficient to explain the increase of $T$ and the observed maximum at higher $f$, which will be discussed in more detail in the text. Let us move on to the case where there are only three layers of tissue.
	
	For the three-layer scenario, we either have a dielectric-conductor-dielectric  (dura-CSF-WM) or a dielectric-conductor-conductor (dura-CSF-GM) interface-- so no capacitor-like configuration. In both cases, a reflection is occuring and a $T$ value that is determined with $f$. This can be attributed to interference with reflected wave and tissue absorption. Hence, we need to verify the characteristic impedance of the tissues to know if the reflections at the junction are hard or soft. Using (\ref{eq:impedance}) and the fact that $\tilde{n}$ of CSF is greater than those of WM and GM, one can conclude that the impedances of WM and GM will also be greater than the one of CSF ($\eta_2 > \eta_1$). This results in soft reflection, meaning that there is no phase shift for the reflected wave, in addition to generating standing waves in the CSF and weaker wave transmission of the EF to GM and/or WM. As the impedance of the CSF and other tissues gets closer to one another with rising $f$, there is a lowering the impedance mismatch between tissues and therefore an increase of $T$ is observed. The same point can be made for the four-layer case. 
	
	In the same order of idea, the $T$ of path c (four layers, thin WM) is close to that of path b (three layers-- GM). This results from the fact that WM ($< 2 \text{ mm}$) is small compared to $\lambda$ ($\sim 0.9 \text{ m}$) of the incident pulse from the CSF. Additionally, it means that the reflection is soft, hence no phase change for the reflected wave. 
	Also, since the WM is so thin, there is little to no absorption of the EF occuring in this frequency regime by the WM according to the Beers-Lambert law of absorption, resulting in a similar $T$ of the path c compared to the one of the bath b.
	
	
	However, these observations alone are insufficient to explain $T$ in the radio frequency (RF) region, nor its rise to a maximum at $\sim 1 \freq{G}$, which is discussed next. 
	
	\subsection{Transmission through multiple layers, high frequency regime} 
	
	For both the three- or four-layer paths, the slope of $T$ has a stronger increase starting from $f \sim 0.16 \freq{G}$ until it reaches a maximum at $f \sim 1 \freq{G}$. The peak at $\sim 1 \freq{G}$ is observed for the different configurations. It is near this $f$ that the CSF behavior transitions from conductor to dielectric, effectively making the spinal cord a four-layer dielectric medium. In addition, it is where the CSF $n''$ drops. This reduces impedance mismatch between tissues and consequently increases $T$. It is consistent with $\varepsilon_r'$, which hits a plateau, and $\varepsilon_r''$ that drops for all tissues, with the exception of the dura, indicating a reduction in tissue absorption of EF energy. This would explain why the $T$ maximum for EF can reach 51\% for a long path (d in figure \ref{FIG6}) and up to 75\% for other configurations (a, b and c in figure \ref{FIG6}) at the discussed frequency. 
	
	From $1 \freq{G}$ onward, $T$ of all paths drops to near zero for $f$ going from $ \sim 10 \freq{G}$ up to $ \sim 30 \freq{G}$. All then stays linear up to $0.1 \freq{T}$, which is the opposite behavior of $\varepsilon_r''$ of the tissues. This is in agreement with the observed transmission rate of dura and CSF of previous work \cite{Chakarothai2021}. It is also where the water dampening effect starts to increase and present molecular viscosity.
	The following broad region ($0.1 \freq{T}$ to $10 \freq{T}$), also known as the ``terahertz gap'', coincides with the expected absorption of EF energy by biological tissue \cite{Frohlich1975,Prohofsky2004,Chakarothai2021,Yada2008,Sheppard2008,Mumtaz2022,Ma2024}. In other words, the drop in $T$ can be attributed in part to water absorption \cite{Prohofsky2004} and tissue composition, as discussed earlier. The water molecule has a rotational resonance at $2.45 \freq{G}$ and exhibits other significant maxima up to $300 \freq{G}$, although weaker the closer $f$ is to $300 \freq{G}$, with an absorption resonance at $\sim 25 \freq{G}$ \cite{Marechal2011,Yada2008,Sheppard2008}. Then, starting at $f \sim 400 \freq{G}$, molecules such as proteins, lipids, and enzymes in the different tissues are not as overdamped by water as before due to the movement of hydrogen bond. They then can absorb the EF energy \cite{Lundholm2015}, maintaining the weak propagation of EF through tissues.
	
	\subsection{Tissue thickness dependency}
	
	Another important factor in the model that has not yet been thoroughly discussed is the thickness of the tissues. We can see that for EF with lower $f$ values, tissue thickness has little to no impact on $T$, as $\lambda$ is multiple order of magnitudes greater than the spinal cord canal. However, starting at frequencies in the range of $10 \freq{M}$, we can see that the medium thickness starts to show a significant impact on $T$. For a total path of $\sim4\text{ mm}$, we see a greater $T$ compared to the one of a longer path of $\sim 7.3 \text{ mm}$ for the mentioned $f$ values. This behavior is in agreement with Beer-Lambert's law, which states that $T$ follows a negative exponential behavior that is driven by to the thickness of the medium and its absorbance coefficient ($T \propto$ exp$\lbrace - \alpha r_m \rbrace$, where $\alpha$ is the absorbtion coefficient and $r_m$ the distance traveled by the EF in a given medium). This also agrees with previous observations \cite{DePaoli2020,Chakarothai2021}.  Finally, as $f$ increases, the tissue thickness gets closer to $\lambda$/2, resulting in destructive interference between the reflected wave and incident EF as well. 
	
	\subsection{Comparison with previous work}
	
	The behavior of $T$ according to $f$ observed here for four layers of tissue is similar to previous work using finite element calculations \cite{Chakarothai2021}, which studies the propagation of EF through six layers of the head for a frequency range ranging from $50 \freq{M}$ to $10 \freq{G}$. This work $T$ peaks at around $1 \freq{G}$ compared to previous work where $T$ peaks near $2 \freq{G}$ with similar values and trend for $T$ (CSF to dura mater). We want to underline that our work studies the propagation of EF through two up to four layers of tissues that make up the spinal cord, starting with the dura, with a frequency range going from $10 \freq{}$ to $0.1 \freq{T}$ to ensure coverage of pulse parameters used clinically. Finally, the spinal cord has a cylindrical geometry. In contrast, previous work focuses on six layers of the head, starting with skin, has tissue thickness that differs from those of the spinal cord and do not include WM. In addition, unlike the cylindrical spine, the head presents more of a spherical geometry, which has a great influence on the propagation of EF. These differences can explain the mismatch in the calculated values of our work compared to other work. However, the results presented are in line with simulations made either with commercial software such as Comsol$^{\text{TM}}$ or simulations coded by research teams \cite{Wesselink1998,Lak2012,Arle2013,Huang2014,Lempka2015,DePaoli2020}.

	\subsection{Limitations and perspectives}
	
	Beyond these results, it must be remembered that this work studies the transmission of a normally incident EF at junctions of different tissues to better our theoretical understanding of what is happening in the spinal cord. In that sense, it presents the upper theoretical limit of $T$ that can be achieved. It is also good to reiterate that the validity of the model relies on a scarce number of experimental results below $f \sim 400 \freq{M}$ \cite{GabrielS1996b,GabrielS1996a,GabrielC1996,Penfold1929,Jacques2013,Zhu2018,Nagel2018,Shapey2021} and that more measurements of the parameters composing $T$  for frequencies below $400 \freq{M}$ are required to further validate the calculated values. Additionally, more theoretical work is required to simulate and explore how the EF propagates within spinal cord tissues using a 3D framework. The current description ($wp$ model) neglects 3D geometrical effects such as Mie scattering and shunting, leading to an overestimation of $T$. Indeed, the strict assumption of a perfectly perpendicular EF polarization propagating in the spinal cord mathematically eliminates any geometrical influence in the presented model. However, real  clinical conditions will present edge effects from the electrodes and three dimensional divergence of the EF in the cylindrical geometry of the spinal cord, resulting in oblique induction. In turn, these components will trigger longitudinal shunting effect in CSF and along axon tracks, altering the distribution of energy reaching the deeper structures, all of which only a 3D model can accurately capture.
	
	As for future work, the use of a 3D framework would allow us to better simulate EF propagation in the spinal cord. In addition, it would be interesting to further explore the impact of different underlying variables composing a pulse such as pwd,  waveform (e.g. square pulse, sinusoidal wave, triangular pulse, etc.), electrode pair configuration and the geometry in which the EF is propagating, as it would change the interactions of the EF with the spinal cord tissues. However, work on the topic is scarce due to technical limitations, such as generating square pulses at $f \geq 30 \freq{k}$, but is indicative that there is great potential to improve SCS efficiency with enhanced stimulation devices \cite{McIntyre2002,Gerasimenko2010,Astrom2012,Huang2014,DePaoli2020,Lempka2015,Canna2021,Rogers2024,Bo2017,Bo2020,Bo2021,Rowald2022}. If a stimulation device capable of producing pulses with $f$ in the GHz is developed, the theoretical results of this work could be used with it to explore the aforementioned parameters. 
		
	\section{Conclusion}
	
	In summary, we present physical models to calculate different tissue characteristics ($\tilde{\varepsilon}_r$, $\tilde{k}$, and $\tilde{n}$) and dual-frequency framework ($QS$ and $wp$) to calculate the transmission coefficient $T$ in the context of spinal cord stimulation. We performed numerical calculations of the different physical properties of tissues within the spinal canal to better understand their response to an applied EF according to the frequency of said EF. Using these, we evaluated the transmission coefficient $T$ of a propagating EF, using quasi-static and sinusoidal description, in the vicinity of the spinal cord for a spectrum range going from $10 \freq{}$ up to $0.1 \freq{T}$. Electrodes used as reference are parallel to the plane of incidence, and different tissue thickness are studied, which is yet to be reported in the known literature. 
	
	Regarding the transmission of the square biphasic pulse with a frequency of $1.6 \freq{k}$, the EF energy transmitted to WM or GM is less than $\sim 3.6 \text{ \%}$. This has two underlying implications. First, the majority of the energy is absorbed in the dura and CSF ($> 65$\%), mostly via induced current and shunting effect. Secondly, we can postulate that it is the body floating in the CSF, rather than the spinal cord’s WM and GM, that are the first structures activated by the EF and driving the physiological response. These results are in agreement with the finite element method and the quasi static EF approach, such as the one used in the literature and by simulation softwares such as Comsol$^{\text{TM}}$, with some discrepancy in the calculated values. These differences between our results and simulation software arise from the fact that we use a theoretical model instead of a semi-hybrid model and the use of a continuous calculation approach instead of a finite element approach.
	
	Finally, the different demeanors observed in $T$ were explained by taking into account the physical and physiological characteristics of the biological tissue that are related to the water composition of the said tissues. We discussed the impact of tissue thickness and how the bulk modes of vibration combined with destructive interference of EF due to tissue thickness suggest that an EF under certain frequency thresholds will be greatly absorbed in the CSF and dura mater. The tissue thickness of WM and GM has little influence on $T$ in the low- and intermediate- frequencies range ($f < 10 \freq{M}$), which contrasts with higher frequency regimes.
	
	In light of these results, we believe that the physiological response to spinal cord stimulation results first from the activation of bodies floating within the CSF, or from the activation of the dorsal horn in a very localized region of the spinal cord where the interposition of WM with the CSF would be minimal before the activation of WM and/or GM. However, due to the low number of measurements of $\tilde{\varepsilon}_r$ and $\tilde{n}$ real and complex parts, further experimental results are required to validate the calculated values and confirm these observations.
	
	\section*{Data availability statement}
	
	In compliance with EPSRC policy framework on research data, this publication is theoretical work that does not require supporting research data.
	
	\section{Acknowledgement}
	This work was supported by the Conseil de recherches en sciences naturelles et en génie du Canada (CRSNG) under the Discovery Horizon grant, file no DH-2022-00922. The authors thank Sarah Labbé, and Dr René Côté for theoretical assistance in the definition of key elements of the electric field modelization.
	
	\section*{Conflict of interest}
	\noindent The authors declare no conflict of interest. \\
	%

\bibliographystyle{apsrev4-2}
\bibliography{PREPRINT_Article01TransmissionOfElectriFieldSpineAPSOS}

\clearpage

\input{PREPRINT_Article01TransmissionOfElectriFieldSpineAPSOSFullCalculation.tex}

\end{document}

%% file: PREPRINT_Article01TransmissionOfElectriFieldSpineAPSOSFullCalculation.tex
\appendix
\section{Supplementary parameters}

\begin{table}[ht]
\caption{\label{table:Param2}%
Human spinal cord distance between electrodes and target tissue. Dimensions taken at vertebra L4 \cite{Hong2011,Papinutto2015,Park2016,Calabres2018,Kinaci2020,Toossi2021} and conductivity values are taken from previous work \cite{Wesselink1998,Capogrosso2013,Seo2015,Zander2020}.}
\begin{ruledtabular}
\begin{tabular}{lccd}
\textrm{Tissue}& \textrm{Thickness (mm)} & Conductivity (S/m)\\
\colrule
 \textrm{GM} & 1 (min) to 5.2 (max) & 0.23 (avg) \\
\textrm{WM} &  1.2 (min) to 4.4 (max) & 0.083 (perp) to 0.6 (para) \\
\textrm{CSF} & 2.38 (avg) & 1.65 \\
\textrm{Dura} & 0.44 (avg) & 0.03 (lower lim) \\
\end{tabular}
\end{ruledtabular}
\end{table}

\section{Detailed calculations}
\setcounter{page}{1} 
\subsection{Refractive index equation}
\label{sec:eqrefr}

If permittivity $\varepsilon$ and conductivity $\sigma$ were real and complex $\tilde{k}$, we could use the approach for conductors where we'd have $-\tilde{k}^2 +\varepsilon \mu \omega^2 +\rmi \omega \mu \sigma =0 $ (or dielectrics using $\rmi \omega \sigma \mu - \omega^2 \varepsilon \mu =\gamma^2$) which would yield\cite{Griffith2013,Khalid2015}: 
\begin{equation}
    \underbrace{k= \beta =  \left\lbrack \frac{1}{2} \omega^2 \varepsilon \mu \left( \sqrt{1 + \left( \frac{\sigma}{\varepsilon \omega} \right)^2 } +1 \right)  \right\rbrack^{1/2} }_{\text{Phase}} , 
\end{equation}
\begin{equation}
    \underbrace{\kappa = \alpha = \left\lbrack \frac{1}{2} \omega^2 \varepsilon \mu \left( \sqrt{1 + \left(\frac{\sigma}{\varepsilon \omega}\right)^2 }  -1 \right) \right\rbrack^{1/2}}_{\text{Attenuation}}.
\end{equation}
All of this holds true for some materials, but not for the biological tissues of interest here. Instead, we have a complex $\tilde{\varepsilon}$. The complex refractive index such medium can be written as:
\begin{equation}
\tilde{n}^2 =  \tilde{\varepsilon}_r .
\end{equation}
With $\tilde{\varepsilon}_r$ complex as well we can write:
\begin{equation}
(n' + \rmi n'')^2 = \varepsilon_r' - \rmi \varepsilon_r', \\
n'^{ 2} - n''^2 + 2 \rmi n' n'' =  \varepsilon_r' + \rmi \varepsilon_r'' .
\end{equation}
Splitting in the real and imaginary parts:
\begin{gather}
\underbrace{n'^{ 2} - n''^2}_{\text{Re}}  = \varepsilon_r' \quad ; \quad \underbrace{2 n' n'' }_{\text{Im}}= \varepsilon_r'' .
\end{gather}
And doing the norm of the refractive index
\begin{gather}
|\tilde{n}^2| = \sqrt{\text{Re}^2 + \text{Im}^2} = \sqrt{(n'^{ 2} - n''^2 )^2 +(2 n' n'')^2 }, \\
\sqrt{(n'^2 + n''^2 )^2} =\sqrt{ \varepsilon_r'^2  +  \varepsilon_r''^2  } ,
\end{gather}
which leads us to:
\begin{eqnarray}
n'^2 + n''^2 =\sqrt{ \varepsilon_r'^2  +  \varepsilon_r''^2  } \quad ; \quad n'^{ 2} - n''^2  = \varepsilon_r'  .
\end{eqnarray}
The sum of the two equations:
\begin{eqnarray}
2 n'^2 = \sqrt{ \varepsilon_r'^2  +  \varepsilon_r''^2  } +  \varepsilon_r'  , \\
n' = \sqrt{\frac{1}{2} \left(\sqrt{ \varepsilon_r'^2  +  \varepsilon_r''^2  } +  \varepsilon_r' \right)   } .
\end{eqnarray}
The difference of the two equations:
\begin{eqnarray}
2 n''^2  = \sqrt{ \varepsilon_r'^2  +  \varepsilon_r''^2  } -  \varepsilon_r' , \\
n''= \sqrt{\frac{1}{2} \left(\sqrt{ \varepsilon_r'^2  +  \varepsilon_r''^2  } -  \varepsilon_r' \right)   }.
\end{eqnarray}
Finally, one can link the phase vector to the refractive index using $\tilde{n} = c \tilde{k} / (2\pi f)$. 

\subsection{Quasi-static model}
\label{sec:QSregimeSM}

In the low frequency region (below $1.6 \freq{M}$), the EF has a wavelength multiple orders of magnitude greater than the size of the entire geometry of interest without and within the mediums. This implies that traveling wave fronts, phase boundaries, optical reflections and refractive indices do not exist inside the spinal canal. For these reasons, we must start to study the EF propagation within the spinal canal using Laplace volume conduction equation:
\begin{gather}
	\nabla^2 \phi = 0 ,
\end{gather}
which has a solution of the form:
\begin{gather}
	\phi = A x + B ,
\end{gather}
with $A_i = E_i $. To further the use of this solution, we can use the continuity of potential at junctions
\begin{eqnarray}
	\phi_i (r_m) = \phi_j (r_m) , \\
	A_i r_m + B_i = A_j r_m + B_j ,
	\label{eq:potjunc}
\end{eqnarray}
and the continuity of charge as well:
\begin{eqnarray}
	\mathbf{J}_i (r_m) = \mathbf{J}_j (r_m) ,
\end{eqnarray}
that can be written as:
\begin{eqnarray}
	\sigma_i A_i = \sigma_j A_j ,\\
	A_i = \frac{\sigma_j }{\sigma_i } A_j .
	\label{eq:APOT}
\end{eqnarray}
We can substitute this relation in (\ref{eq:potjunc}) and isolate $B_i$:
\begin{eqnarray}
	 \frac{\sigma_j }{\sigma_i } A_j  r_m + B_i = A_j r_m + B_j , \\
	B_i = r_m \left(1 - \frac{\sigma_j }{\sigma_i } \right) A_j + B_j .
	\label{eq:potmatr}
\end{eqnarray}
Using (\ref{eq:APOT}) and (\ref{eq:potmatr}), we can write the following relation for a given junction:
\begin{gather}
\begin{bmatrix}
		A _i \\
		B_i
	\end{bmatrix} 
	=
	C_i
	\begin{bmatrix}
		A _j \\
		B_j
	\end{bmatrix} , \\
	\begin{bmatrix}
		A _i \\
		B_i
	\end{bmatrix}
	=
	\begin{bmatrix}
	\sigma_j / \sigma _i & 0 \\
	r_m (1 - \sigma_j/\sigma_i ) & 1
	\end{bmatrix}
	\begin{bmatrix}
		A_j \\
		B_j 
	\end{bmatrix} .
\end{gather}
Generalizing for three junctions (four tissues), we get:
\begin{gather}
	\begin{bmatrix}
		A_1 \\
		B_1
	\end{bmatrix}
	=
	C_1 C_2 C_3
	\begin{bmatrix}
		A_4 \\
		B_4
	\end{bmatrix} .
\end{gather}
The full matrix multiplication yields:
\begin{gather}
	\begin{bmatrix}
		A _1 \\
		B_1
	\end{bmatrix}
	= \mathbf{D} 
	 \begin{bmatrix} A_4 \\
	 	B_4 
        \end{bmatrix} = 
	\begin{bmatrix}
	\sigma_4 / \sigma _1 & 0 \\
	r_1 \left(\frac{\sigma_4 }{\sigma_2} - \frac{\sigma_4 }{\sigma_1 }\right)  + r_2  \left(\frac{\sigma_4}{\sigma_3} - \frac{\sigma_4 }{\sigma_2 } \right) + r_3  \left(1 - \frac{\sigma_4 }{\sigma_3 } \right) & 1
	\end{bmatrix}
	\begin{bmatrix}
		A_4 \\
		B_4 
	\end{bmatrix} .
	\label{eq:25matrix}
\end{gather}

Now, posing the initial potential at the junction between the electrodes and dura mater as $\phi_0 (r_0 =0) = V_0' $, we get:
\begin{gather}
	\phi_0 = V_0' = A_1 r_0 + B_1 , \\
	V_0' = B_1 .
\end{gather}
Instinctively, one would use this to isolate $A_4$ (at the target tissue) and get $T = |\sigma_1 / \sigma_4|^2 $, which ignores the effect of intermediate layers. To solve this issue, we have to look at the junction between dura mater and CSF where we have:
\begin{gather}
	\phi (r_1) = V_0 =  A_1 r_1 + B_1 ,
\end{gather} 
from which we can isolate $A_1$:
\begin{gather}
	A_1  = \frac{V_0 - V_0' }{ r_1} .
	\label{eq:potdura}
\end{gather}
Using (\ref{eq:25matrix}), and the fact that the potential in deep tissues is null, hence $B_4 =0$, we get:
\begin{gather}
	A_1 = D_{00} A_4 , \\
	B_1 = D_{10} A_4 .
\end{gather}
Substituting those in (\ref{eq:potdura}), one can write:
\begin{gather}
	D_{00} A_4 = \frac{V_0 - D_{10} A_4 }{r_1} ,\\
	A_4 ( r_1 D_{00} + D_{10}) = V_0,
\end{gather}
and finally, we can get
\begin{gather}
	E_4 = A_4 = \frac{V_0 }{ r_1 D_{00} + D_{10} } .
\end{gather}
Knowing that $E_{source} = V_0 / r_1$, the reference EF scaled to the first junction to ensure that $T$ accounts for the effects of intermediate layers, we can write
\begin{gather}
	T = \left| \frac{E_4 }{E_{source}} \right|^2  = \left| \frac{ r_1 }{ r_1 D_{00} +D_{10} } \right|^2 .
	\label{eq:TQS4}
\end{gather}
The same approach can be used for two junctions (three layers) of tissues, which will yield the same equation of $T$ as \ref{eq:TQS4}, but with $D_{00}$ and $D_{10}$ written as
\begin{gather}
	D_{00} = \frac{\sigma_3 }{\sigma_1 } , \\
	D_{10} = r_1 \left( \frac{\sigma_3}{\sigma_2 } - \frac{\sigma_3}{\sigma_1} \right) + r_2 \left( 1 - \frac{\sigma_3}{\sigma_2} \right) .
\end{gather}

\subsection{Wave propagation model}
\label{sec:AppCalculs}

When the EF wavelength starts to be closer to the mediums dimension, we can use Helmholtz wave propagation approach. At first, we assume a normal incident EF that is perfectly perpendicular to WM fibers. We also assume that the different tissues of the spine behave like lossy dielectrics. We then described the junctions as a continuation of the electric field with incident, transmitted, and reflected parts ($E_I + E_R = E_T$) with exponentials (the temporal part of the exponentials cancels out each other). Using the complex form of the wave vector $\tilde{k}$, which holds for lossy mediums as the lossy effects are within $k$ and $\kappa$,  it is possible to write (to lighten the writing, we'll discard the complex notation ($\sim$) of the refractive index $n$) \cite{Borne2002,Griffith2013}:
\begin{gather}
	E^{(j)}_\RR \rme^{\rmi k_j r_j -\kappa_j r_j  } + E^{(j)}_\LL \rme^{-\rmi k_j r_j +\kappa_j r_j }   = E^{(j+1)}_\RR e^{\rmi k_{j+1} r_j -\kappa_{j+1} r_j } + E^{(j+1)}_\LL \rme^{-i k_{j+1} r_j +\kappa_{j+1} r_j}, \label{eq:condfrontfirst} 
\end{gather}
\begin{gather}
	 E^{(j)}_\RR \rme^{i k_j r_j -\kappa_j r_j } -  E^{(j)}_\LL \rme^{-\rmi k_j r_j +\kappa_j r_j }    = \frac{n_{j+1} }{n_j} {\Big (} E^{(j+1)}_\RR \rme^{\rmi k_{j+1} r_j -\kappa_{j+1} r_j } - E^{(j+1)}_\LL \rme^{-\rmi k_{j+1} r_j +\kappa_{j+1} r_j } {\Big )}. 
\end{gather}
Using this definition, we can write the junction equations. At the first junction, Dura to CSF (z=0) (1 to 2):
\begin{eqnarray}
    E_\RR^{(1)} + E_\LL^{(1)} &= E_\RR^{(2)} + E_\LL^{(2),}  \\
    E_\RR^{(1)} -E_\LL^{(1)} &= \frac{n_2}{n_1} (E_\RR^{(2)}-E_\LL^{(2)}) ,
    \label{eq:milieuelecdebut}
\end{eqnarray}
 CSF to WM or GM (z=a) (2 to 3):
\begin{gather}
  E_\RR^{(2)} \rme^{(\rmi k_2  - \kappa_2) a} +E_\LL^{(2)}\rme^{-(\rmi k_2 - \kappa_2) a} 
  = E_\RR^{(3)} \rme^{(\rmi k_3 -\kappa_3) a} +E_\LL^{(3)} \rme^{-(\rmi k_3 -\kappa_3) a},  
  \end{gather}
  \begin{gather}
   E_\RR^{(2)} \rme^{(\rmi k_2  - \kappa_2) a} -E_\LL^{(2)} \rme^{-(\rmi k_2 - \kappa_2) a}  =\frac{n_3}{n_2} \left( E_\RR^{(3)}\rme^{(\rmi k_3 -\kappa_3) a} -E_\LL^{(3)} \rme^{-(\rmi k_3 -\kappa_3) a} \right).
\end{gather}
Junction WM to GM (z=b) (3 to 4):
\begin{gather}
   E_\RR^{(3)} \rme^{(ik_3  - \kappa_3) b} +E_\LL^{(3)} \rme^{-(\rmi k_3 - \kappa_3) b}   = E_\RR^{(4)} \rme^{( \rmi k_4 -\kappa_4) b} +E_\LL^{(4)} \rme^{-(\rmi k_4 -\kappa_4) b},
   \end{gather}
   \begin{gather}
   E_\RR^{(3)} \rme^{(\rmi k_3  - \kappa_3) b}  -E_\LL^{(3)} \rme^{-(\rmi k_3 - \kappa_3) b} =\frac{n_4}{n_3} \left( E_\RR^{(4)} \rme^{(\rmi k_4 -\kappa_4) b} -E_\LL^{(4)}\rme^{-(\rmi k_4 -\kappa_4) b} \right). 
    \label{eq:milieuelecfin}
\end{gather}
These equations describe an EF with normal incidence.  Substituting equations for a matrix system of the form $M_{ij} E_i = N_{kl}E_{k}$, one can write the system of transfer matrix as \cite{Balili2012}:
\begin{gather}
    M_1 E_1 = N_1 E_2 \rightarrow
    E_1 = M_1^{-1} N_1  E_2  ,\\
    M_2 E_2 = N_2 E_3 \rightarrow
    E_2 = M_2^{-1} N_2   E_3,\\
    M_3 E_3 = N_3 E_4 \rightarrow E_3 =M_3^{-1} N_3 E_4, \\
    E_1 = M_1^{-1} N_1 M_2^{-1} N_2 M_3^{-1} N_3 E_4 =\mathbf{F} E_4, 
    \label{eq:finalind}
\end{gather}
and we can rewrite $\mathbf{F}$ using index notation as:
\begin{eqnarray}
     A_{ik} B_{kl} C_{lm} D_{mn} G_{no} H_{oj} = F_{ij}.
    \label{eq:matrEFieldind}
\end{eqnarray}
Since the transmission coefficient is the ratio between the transmitted and incident EF amplitudes, it can be written as
\begin{equation}
	T =  \frac{I_\TT}{I_0} =  \frac{n_\TT }{n_\rmi} \left| \frac{E_\TT }{E_\rmi} \right|^2 ,
\end{equation}
\begin{equation}
	T = \frac{n_4 }{ n_1} \left| \frac{E_4 }{E_1 } \right|^2  = \frac{n_4}{n_1} \left| \frac{ E_{\RR}^{(4) } }{ E_{\RR}^{(1)} + E_{\LL}^{(1) }     } \right|^2 .
\end{equation}
Using the relation (\ref{eq:finalind}) 
combined with the previous one, one can find $E_R^{(1)} + E_L^{(1)} = F_{00} E_R^{(4)} + F_{11} E_L^{(4)}$. Since $E_L^{(4)} =0$ and we study a field with normal incidence with the surface, we get:
\begin{eqnarray}
	T = \frac{n_4}{n_1} \left|\frac{1}{F_{00} }\right|^2 , \\
	=  \frac{n_4}{n_1} \frac{1}{ (\text{Re}\lbrace F_{00} \rbrace)^2 +(\text{Im}\lbrace F_{00} \rbrace)^2  }.
\end{eqnarray}
Which justifies working only with $F_{00}$, and we can write
\begin{eqnarray}
	F_{00}= A_{0k}B_{kl} C_{lm} D_{mn} G_{no} H_{o0}.
\end{eqnarray}
We first define
	\begin{eqnarray}
	D_{00} (G_{00} H_{00} + G_{01} H_{10}) + D_{01} (G_{10} H_{00} + G_{11} H_{10})  = \alpha \nonumber \\ 
	D_{10} (G_{00} H_{00} + G_{01} H_{10}) + D_{11} (G_{10} H_{00} + G_{11} H_{10})   = \beta \nonumber 
\end{eqnarray}
we then get:
\begin{gather}
	F_{00} = A_{00} \lbrace  B_{00} \lbrack C_{00} \alpha + C_{01} \beta \rbrack + B_{01} \lbrack C_{10} \alpha + C_{11} \beta \rbrack \rbrace \nonumber +A_{01} \lbrace   B_{10} \lbrack C_{00} \alpha + C_{01} \beta \rbrack + B_{11} \lbrack C_{10} \alpha + C_{11} \beta \rbrack  \rbrace .
	\label{eq:indicielini}
\end{gather}
We can now write the matrices for the different boundary condition.Using the fact that the WM is anisotropic and the fibers will act as an analyzer, let's begin with the case where the EF polarization is perpendicular to such fibers, simplifying the writing at first.The matrices are written (at z=0):
\begin{eqnarray}
    \begin{bmatrix}
        1 & 1 \\
        1 & -1
    \end{bmatrix}
    \begin{bmatrix}
        E_\RR^{(1)} \\ E_\LL^{(1)}
    \end{bmatrix}
    &= \begin{bmatrix}
        1 & 1 \\
        \frac{n_2}{n_1} & - \frac{n_2}{n_1}
    \end{bmatrix}
    \begin{bmatrix}
        E_\RR^{(2)} \\ E_\LL^{(2)}
    \end{bmatrix},
\end{eqnarray}

and for following junctions:

\begin{gather}
   \begin{bmatrix}
        \rme^{ (\rmi k_{k}-\kappa_k) r_{k}} & \rme^{-(\rmi k_k -\kappa_k) r_k} \\
        \rme^{(\rmi k_{k}-\kappa_k) r_{k}} & -\rme^{-(\rmi k_k -\kappa_k) r_k}
    \end{bmatrix}
    \begin{bmatrix}
        E_\RR^{(k)} \\ E_\LL^{(k)}
    \end{bmatrix} 
     = \begin{bmatrix}
       \rme^{ (\rmi k_{l}-\kappa_l) r_{k}} & \rme^{-(\rmi k_l -\kappa_l) r_k} \\
        \frac{n_l}{n_k} \rme^{(i k_{l}-\kappa_l) r_{k}} & - \frac{n_l}{n_k} \rme^{-(ik_l -\kappa_l) r_k}
    \end{bmatrix}
    \begin{bmatrix}
        E_\RR^{(l)} \\ E_\LL^{(l)}
    \end{bmatrix}.
\end{gather}
Since we want to know what reaches the GM at the last junciton and not what propagates through it, the $\kappa_4$ elements will be neglected in the matrices presented below. The complete system is then written:
 \begin{widetext}
\begin{multline}
    \begin{bmatrix}
        1 & 1 \\
        1 & -1
    \end{bmatrix}^{-1}
    \begin{bmatrix}
        1 & 1 \\
        \frac{n_2}{n_1} & - \frac{n_2}{n_1}
    \end{bmatrix}
    \begin{bmatrix}
        \rme^{( \rmi k_{2}-\kappa_2) a} & \rme^{-(\rmi k_{2}-\kappa_2)  a} \\
        \rme^{(\rmi k_{2}-\kappa_2) a} & -\rme^{-(\rmi k_{2}-\kappa_2)  a}
    \end{bmatrix}^{-1} 
    \begin{bmatrix}
        \rme^{(\rmi k_{3}-\kappa_3)  a} & \rme^{-(\rmi k_{3}-\kappa_3)  a} \\
        \frac{n_3}{n_2}\rme^{(\rmi k_{3}-\kappa_3)  a} & - \frac{n_3}{n_2}  \rme^{-(\rmi k_{3}-\kappa_3)  a}
    \end{bmatrix} \\
   \begin{bmatrix}
        \rme^{(\rmi k_{3}-\kappa_3) b} & \rme^{-( \rmi k_{3}-\kappa_3)  b} \\
        \rme^{(\rmi k_{3}-\kappa_3) b} & -\rme^{-(\rmi k_{3}-\kappa_3)  b}
    \end{bmatrix}^{-1} 
    \begin{bmatrix}
        \rme^{\rmi k_{4}  b} & \rme^{-\rmi k_{4} b} \\
        \frac{n_4}{n_3}\rme^{\rmi k_{4}b} & - \frac{n_4}{n_3}  \rme^{-\rmi k_{4} b}
    \end{bmatrix}
     E_4 = E_1 ,
\end{multline}
\end{widetext}
which leads to
\begin{widetext}
\begin{multline}
     \frac{1}{8}
    \begin{bmatrix}
        1 & 1 \\
        1 & -1
    \end{bmatrix}
    \begin{bmatrix}
        1 & 1 \\
        \frac{n_2}{n_1} & - \frac{n_2}{n_1}
    \end{bmatrix}
    \begin{bmatrix}
        \rme^{-(\rmi k_{2}-\kappa_2) a} & \rme^{-(\rmi k_{2}-\kappa_2) a} \\
        \rme^{(\rmi k_{2}-\kappa_2)a} & -\rme^{(\rmi k_{2}-\kappa_2) a}
    \end{bmatrix} 
    \begin{bmatrix}
        \rme^{(\rmi k_{3}-\kappa_3) a} &\rme^{-(\rmi k_{3}-\kappa_3) a} \\
        \frac{n_3}{n_2} \rme^{(\rmi k_{3}-\kappa_3) a} & - \frac{n_3}{n_2}  \rme^{-(\rmi k_{3}-\kappa_3) a}
    \end{bmatrix}  \\
 \begin{bmatrix}
        \rme^{-(\rmi k_{3}-\kappa_3)  b} & \rme^{-(\rmi k_{3}-\kappa_3)  b} \\
        \rme^{(\rmi k_{3}-\kappa_3)  b} & -\rme^{(\rmi k_{3}-\kappa_3)  b}
    \end{bmatrix} 
    \begin{bmatrix}
        \rme^{ \rmi k_{4} b} & \rme^{- \rmi k_{4}  b} \\
        \frac{n_4}{n_3} \rme^{ \rmi k_{4} b} & - \frac{n_4}{n_3}  \rme^{- \rmi k_{4} b}
    \end{bmatrix}
     E_4 = E_1.
\end{multline}
\end{widetext}

Reminding the following  exponential relations:
\begin{eqnarray}	
	\rme^{\rmi x} = \cos(x) + \rmi \sin(x) \quad ; \quad \rme^{-\rmi x} = \cos(x) - \rmi \sin(x), \nonumber \\
	\rme^{x} + \rme^{-x} = 2 \cosh(x) \quad ; \quad \rme^{x} - \rme^{-x} = 2 \sinh(x). \nonumber
\end{eqnarray}
To lighten the writing of the equations, the following abbreviations are used:
\begin{eqnarray}
	c_i = \cos(k_i (r_i - r_{i+1})) \quad ; \quad s_i = \sin(k_i (r_i - r_{i+1}) )  \nonumber ,\\ 
	ch_i = \cosh(\kappa_i (r_i - r_{i+1})) \quad ; \quad sh_i = \sinh(\kappa_i (r_i - r_{i+1})) , \nonumber 
\end{eqnarray}
as an example:
\begin{eqnarray}
	c_3 = \cos(k_3 (a-b)) \quad ; \quad s_3 = \sin(k_3 (a-b)), \nonumber \\
	ch_3 = \cosh(\kappa_3 (a-b)) \quad ; \quad sh_3 = \sinh(\kappa_3 (a-b)), \nonumber
\end{eqnarray}
with the exception of $i=2$, where instead of $r_i - r_{i+1}$ we only have $a$ as a distance. Using these in equation (\ref{eq:indicielini}) and reminding the factor 1/8 multiplying $\mathbf{F}$, we finally get:
\begin{widetext}
\begin{multline}
F_{00} = \frac{  \rme^{ \rmi k_{4} b} }{2} ch_3 {\Big \lbrack} ch_2 c_2 \left\lbrace c_3 \left(1+\frac{n_4}{n_1} \right) + \rmi s_3 \left( \frac{n_3}{n_1} + \frac{n_4}{n_3} \right) \right\rbrace -\rmi sh_2 s_2 \left\lbrace c_3 \left( 1 +\frac{n_4}{n_1} \right)+ \rmi s_3 \left(\frac{n_3}{n_1} +\frac{n_4}{n_3} \right) \right\rbrace   \\
	 +sh_2 c_2 \left\lbrace c_3 \left(\frac{n_2}{n_1} +\frac{n_4}{n_2} \right) +\rmi s_3 \left( \frac{n_4}{n_3} \frac{n_2}{n_1} + \frac{n_3}{n_2} \right)\right\rbrace - \rmi ch_2 s_2 \left\lbrace c_3 \left(\frac{n_2}{n_1}+\frac{n_4}{n_2} \right) +\rmi s_3 \left(\frac{n_4}{n_3}\frac{n_2}{n_1} +\frac{n_3}{n_2} \right) \right\rbrace {\Big \rbrack}  \\
	 - \frac{  \rme^{ \rmi k_{4} b} }{2} sh_3 {\Big \lbrack} ch_2 c_2 \left\lbrace c_3 \left(\frac{n_4}{n_3} +\frac{n_3}{n_1}\right)+ \rmi s_3 \left(1+ \frac{n_4}{n_1} \right) \right\rbrace -\rmi sh_2s_2 \left\lbrace   c_3 \left( \frac{n_4}{n_3}+\frac{n_3}{n_1}\right) +\rmi s_3 \left( 1+\frac{n_4}{n_1} \right)   \right\rbrace  \\
	+sh_2 c_2 \left\lbrace c_3\left( \frac{n_4}{n_3} \frac{n_2}{n_1} + \frac{n_3}{n_2} \right) + \rmi s_3 \left( \frac{n_2}{n_1} +\frac{n_4}{n_2}\right) \right\rbrace - \rmi ch_2 s_2 \left\lbrace  c_3 \left( \frac{n_4}{n_3}\frac{n_2}{n_1} +\frac{n_3}{n_2} \right) + \rmi s_3 \left( \frac{n_2}{n_1} + \frac{n_4}{n_2} \right)   \right\rbrace {\Big \rbrack}. 
\end{multline}
\end{widetext}
Keeping in mind that $n_{(i \text{ or } j)}$ is complex, the ratio $n_i/n_j$ gives: 
\begin{gather}
	\frac{n_i }{n_j} = \frac{n_i'+ \rmi n_i''}{n_j' + \rmi n_j''} =  \left( \frac{n_i'+ \rmi n_i''}{n_j' + \rmi n_j''} \right) \left( \frac{n_j' - \rmi n_j''}{n_j' - \rmi n_j''}\right) =  \frac{n_i' n_j' + n_i'' n_j'' + \rmi(n_i'' n_j' - n_i' n_j'')}{n_j'^2 + n_j''^2}. \label{eq:ajoutnsurn}
\end{gather}
And the ratio $(n_i/n_j)(n_l/n_m)$ yields:
\begin{widetext}
\begin{multline}
	\frac{ n_i }{n_j} \frac{n_l }{n_m} = \frac{ (n_i' n_j' + n_i'' n_j'') (n_l' n_m' + n_l'' n_m'')-  (n_i'' n_j' - n_i' n_j'')(n_l'' n_m' - n_l' n_m'')}{(n_m'^2 + n_m''^2)(n_j'^2 + n_j''^2)   } \\ + \rmi \left( \frac{(n_l'' n_m' - n_l' n_m'')(n_i' n_j' + n_i'' n_j'') + (n_m' n_l' + n_l'' n_m'') (n_i'' n_j' - n_i' n_j'')   }{(n_m'^2 + n_m''^2)(n_j'^2 + n_j''^2)   } \right)  .	\label{eq:ajoutnnsurnn}
\end{multline}
\end{widetext}
To lighten the writing from now on, we'll use $n_i$$'/n_j $ along $n_i n_l$$' (n_j n_m)^{-1}$ to represent the real part, and $n_i$$'' /n_j$ along $n_i n_l$$'' (n_j n_m)^{-1}$ to represent the complex part of (\ref{eq:ajoutnsurn}) and (\ref{eq:ajoutnnsurnn}) in the following equations. The real part of F$_{00}$ is :
\begin{widetext}
\begin{multline}
\text{Re}\lbrace F_{00} \rbrace=  \frac{   \rme^{ \rmi k_{4} b} }{2} {\Big ( } c_2 ch_2 c_3 \left\lbrack ch_3 \left(1 + \frac{n_4}{n_1}' \right) - sh_3 \left(\frac{n_4}{n_3}' +\frac{n_3}{n_1}' \right)  \right\rbrack + c_2 sh_2 c_3  {\Big \lbrack } ch_3 \left( \frac{n_2}{n_1}' +\frac{n_4}{n_2}' \right)  \\
 - sh_3 \left( \frac{n_4}{n_3} \frac{n_2}{n_1}' + \frac{n_3}{n_2}' \right) {\Big \rbrack } + s_2 sh_2 s_3 \left\lbrack ch_3 \left(\frac{n_3}{n_1}' +\frac{n_4}{n_3}' \right) - sh_3 \left(1+\frac{n_4}{n_1}' \right)\right\rbrack  \\+s_2 ch_2 s_3 { \Big \lbrack} ch_3 \left( \frac{n_4}{n_3} \frac{n_2 }{n_1}' + \frac{n_3}{n_2}' \right)  - sh_3 \left(\frac{n_2}{n_1}' + \frac{n_4}{n_2}' \right) { \Big \rbrack }  +c_2 ch_2 s_3 {\Big \lbrack} -ch_3 \left(\frac{n_3}{n_1}'' + \frac{n_4}{n_3}'' \right) + sh_3 \frac{n_4}{n_1}'' {\Big \rbrack}  \\
+s_2 sh_2 c_3 {\Big \lbrack} ch_3 \frac{n_4 }{n_1}'' - sh_3 \left( \frac{n_4}{n_3}'' +\frac{n_3}{n_1}'' \right) {\Big \rbrack} +c_2 sh_2 s_3 {\Big \lbrack } ch_3 \left(\frac{n_4 n_2}{ n_3 n_1}'' + \frac{n_3}{n_2}'' \right)  \\
+ sh_3 \left(\frac{n_2}{n_1}''+ \frac{n_4}{n_2}'' \right){\Big \rbrack}  +s_2 ch_2 s_3 {\Big \lbrack} ch_3 \left( \frac{n_2}{n_1}'' \right)- sh_3 \left(\frac{n_4 n_2}{ n_3 n_1}'' + \frac{n_3}{n_2}'' \right) {\Big \rbrack}  {\Big )}.
\end{multline}
\end{widetext}
The imaginary part is:
\begin{widetext}
\begin{multline}
	\text{Im}\lbrace F_{00} \rbrace =  \frac{   \rme^{ \rmi k_{4} b} }{2} {\Big (}-  s_2 sh_2 c_3 \left\lbrack ch_3 \left(1+\frac{n_4}{n_1}' \right)-sh_3 \left( \frac{n_4}{n_3}' +\frac{n_3}{n_1}' \right) \right\rbrack  - s_2ch_2 c_3 {\Big \lbrack} ch_3 \left( \frac{n_2}{n_1}' + \frac{n_4}{n_2}' \right)  \\ 
	- sh_3 \left( \frac{n_4}{n_3} \frac{n_2}{n_1}' + \frac{n_3}{n_2}' \right) {\Big \rbrack}  +  c_2 ch_2 s_3 \left\lbrack ch_3 \left(\frac{n_3}{n_1}' + \frac{n_4}{n_3}' \right) -sh_3 \left( 1+ \frac{n_4}{n_1}' \right)\right\rbrack  \\
	 + c_2 sh_2 s_3 \left\lbrack ch_3 \left(\frac{n_4}{n_3} \frac{n_2}{n_1}' + \frac{n_3}{n_2}' \right) - sh_3 \left(\frac{n_2}{n_1}' + \frac{n_4}{n_2}' \right) \right\rbrack  + c_2 ch_2 c_3 {\Big \lbrack } ch_3 \frac{n_4}{n_1}'' -sh_3 {\Big (}\frac{n_4}{n_3}'' + \frac{n_3}{n_1}'' {\Big )} {\Big \rbrack}  \\ 
	 -s_2 sh_2 s_3 {\Big \lbrack } -ch_3  {\Big (} \frac{n_3}{n_1}'' + \frac{n_4}{n_3}''  {\Big )} - sh_3 \frac{n_4}{n_1} ''{\Big \rbrack} + c_2 sh_2 c_3 {\Big \lbrack } ch_3  {\Big (} \frac{n_2}{n_1}'' + \frac{n_4}{n_2}''  {\Big )}  \\
	  -sh_3  {\Big (} \frac{n_4}{n_3}'' \frac{n_2}{n_1} '' +\frac{n_3}{n_2}''  {\Big )} {\Big \rbrack} - s_2 ch_2 s_3 {\Big \lbrack } -ch_3  {\Big (} \frac{n_4}{n_3}'' \frac{n_2}{n_1}''  + \frac{n_3}{n_2}''  {\Big )} +sh_3  {\Big (} \frac{n_2}{n_1}'' + \frac{n_4}{n_2}'' {\Big )}  {\Big \rbrack }  {\Big )}.
	\end{multline}
\end{widetext} 	
\noindent The same exercise can be done for three layers instead of four, using 
\begin{eqnarray}
	A_{ik} B_{kl} C_{lm} D_{mj}  = F_{ij}.
\end{eqnarray}
This will yield the result for $T$ for three junctions (two layers).